\documentclass[usenatbib]{mn2e}

\usepackage{mycite}
\usepackage{amssymb}
\usepackage{amsmath}
\usepackage{graphicx}
\usepackage{natbib}

\newcommand\fsec{\hbox{$.\!\!^{\rm s}$}}

\def\Zsol{\hbox{Z$_{\odot}$}}
\def\Msol{\hbox{M$_{\odot}$}}
\def\one{\,{\sc i}}
\def\two{\,{\sc ii}}
\def\three{\,{\sc iii}}
\def\four{\,{\sc iv}}

\usepackage{epsfig}

\newcommand{\hi}{H\,{\sc i}}
\newcommand{\hii}{H~{\sc ii}}
\newcommand{\hei}{He~{\sc i}}
\newcommand{\heii}{He~{\sc ii}}
\newcommand{\eld}{$N_{\rm e}$}

\newcommand{\elt}{$T_{\rm e}$}

\newcommand{\cmt}{cm$^{-3}$}

\newcommand{\foiii}{[O~{\sc iii}]}

\newcommand{\fsii}{[S~{\sc ii}]}

\newcommand{\fnii}{[N~{\sc ii}]}

\newcommand{\ffeiii}{[Fe~{\sc iii}]}

\newcommand{\niii}{N~{\sc iii}}

\newcommand{\civ}{C~{\sc iv}}

\newcommand{\hb}{H$\beta$}

\title[A VIMOS IFU study of the BCD Mrk~996]{A VLT VIMOS study of the anomalous BCD Mrk~996: mapping the
ionised gas kinematics and abundances$^{\dag}$
\author[B. L. James et al.]
{B.~L. James$^{1}$\thanks{E-mail: bj@star.ucl.ac.uk (BLJ); tsamis@iaa.es
(YGT)}, Y.~G. Tsamis$^{1, 2\,\star}$, M.~J. Barlow$^{1}$, M.~S.
Westmoquette$^1$, J.~R. Walsh$^3$,
\newauthor
F. Cuisinier$^4$, and K.~M. Exter$^5$\\
$^1$Department of Physics and Astronomy, University College London, Gower Street, London, WC1E 6BT\\
$^2$Instituto de Astrof\'{i}sica de Andaluc\'{i}a (CSIC), Apartado 3004, 18080 Granada, Spain\\
$^3$Space Telescope European Co-ordinating Facility, European Southern Observatory, Karl-Schwarzschild Strasse 2, D-85748 Garching, Germany\\
$^4$GEMAC -- Departamento de Astronomia/Observat\'{o}rio do Valongo,
Universidade Federal do Rio de Janeiro, Ladeira do Pedro Antonio 43, 20080-090 Rio de Janeiro RJ, Brazil\\
$^5$ Inst. voor Sterrenkunde, Celestijnenlaan 200D BUS 2401, 3001 Heverlee, Leuven, Belgium\\
\dag{Based on observations made with ESO telescopes at the Paranal Observatory
under programme ID 078.B-0353(A)}}}

\begin{document}

\date{Accepted 2009 May 25.  Received 2009 April 10; in original form 2008 December 18}

\pagerange{\pageref{firstpage}--\pageref{lastpage}} \pubyear{2002}

\maketitle

\label{firstpage}

\begin{abstract}

A study of the blue compact dwarf (BCD) galaxy Mrk~996 based on high resolution
optical VLT VIMOS integral field unit spectroscopy is presented. Mrk~996
displays multi-component line emission, with most line profiles consisting of a
narrow, central Gaussian (FWHM$\sim$110~km s$^{-1}$) with an underlying broad
component (FWHM$\sim$400 km s$^{-1}$). The broad \hi\ Balmer component splits
into two separate broad components inside a 1$''$.5 radius from the nucleus;
these are attributed to a two-armed mini-spiral. This spiral-like nucleus
rotates in the same sense as the extended narrow-line ionised gas but is offset
by $\sim$50 km s$^{-1}$ from the systemic velocity of the galaxy. The rotation
curve of Mrk~996 derived from the H$\alpha$ narrow component yields a total
mass of 5$\times$10$^8$\Msol\ within a radius of 3~kpc. From the H$\alpha$
luminosity we infer a global star formation rate of
$\sim$2\,M$_\odot$~yr$^{-1}$.

The high excitation energy, high critical density \foiii\ $\lambda$4363 and
\fnii\ $\lambda$5755 lines are only detected from the inner region and exist
purely in broad component form, implying unusual excitation conditions. Surface
brightness, radial velocity, and FWHM maps for several emission components are
presented. A separate physical analysis of the broad and narrow emission line
regions is undertaken. We derive an upper limit of 10,000~K for the electron
temperature of the narrow line gas, together with an electron density of 170
cm$^{-3}$, typical of normal \hii\ regions. For the broad line component, 
 measured \foiii\ and \ffeiii\ diagnostic line ratios are consistent 
with a 
temperature of 11,000~K and an electron density of 10$^7$ cm$^{-3}$. 
The broad line emission regions show N/H and N/O enrichment factors of 
$\sim$20 relative to the narrow line regions, but no He/H, O/H, S/H, or 
Ar/H enrichment is inferred.
Previous studies indicated that Mrk~996 showed anomalously high N/O ratios
compared with BCDs of a similar metallicity. Our multi-component analysis
yields a revised metallicity of $\geq$0.5\,\Zsol (12$+$log 
O/H~$=$~8.37) for both the narrow and broad gas components, 
significantly higher than previous studies. As a result the narrow line 
region's N/O ratio is now typical for the galaxy's metallicity.
The narrow line component's N/O ratio peaks outside the core region, spatially correlating
with $\sim$3\,Myr-old stellar populations. The dominant line excitation
mechanism is photoionisation by the $\sim$3000 WR stars and $\sim$150,000
O-type stars estimated to be present in the core. This is indeed a peculiar
BCD, with extremely dense zones of gas in the core, through which stellar
outflows and possible shock fronts permeate contributing to the excitation of
the broad line emission.

\end{abstract}


\section{Introduction}

Blue Compact Dwarf (BCD) galaxies provide a means of studying chemical
evolution and star formation in low metallicity environments in the nearby
Universe.  In general BCDs are faint (M$_{\rm B}$ $>$ 18), have blue optical
colours, are typically small ($\lesssim$1~kpc) and are thought to be
experiencing bursts of star formation in relatively chemically un-evolved
environments, ranging from 1/2--1/50 solar metallicity \citep[][]{Kunth:2000},
and are therefore good analogues to high redshift star forming galaxies.

\begin{figure*}
\begin{center}
\includegraphics[scale=0.70, angle=0]{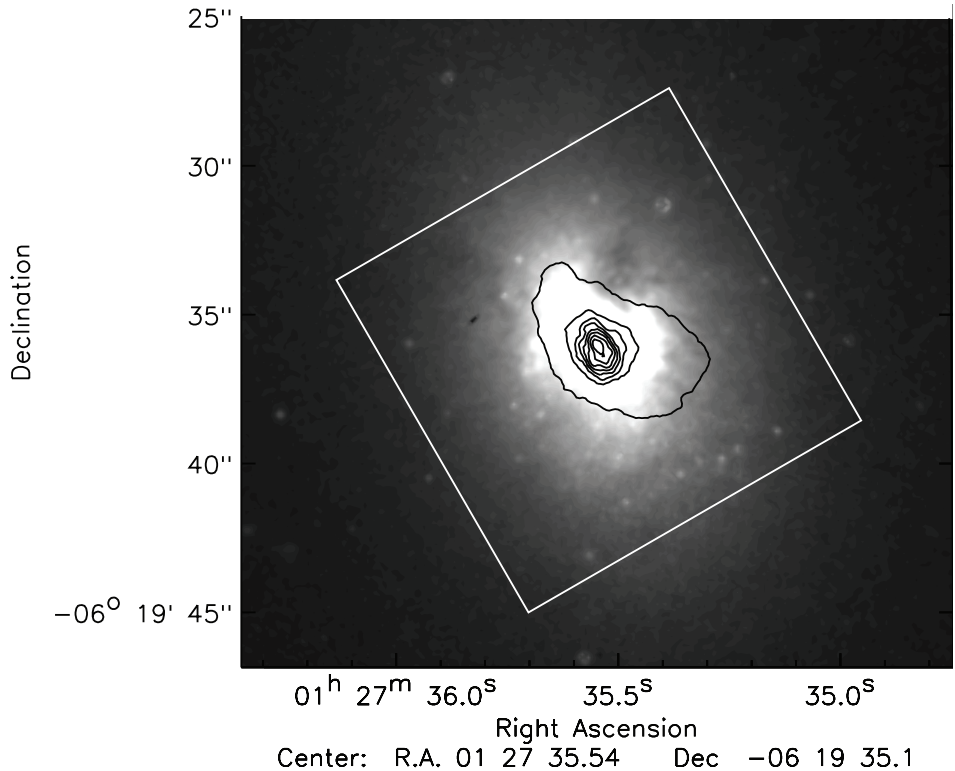}
\caption{{\it HST} F569W WFPC2 image of Mrk~996 overlaid with the 13$''
\times$13$''$ VIMOS IFU aperture at a position angle of 30$^\circ$, along with
contours from the VIMOS IFU H$\alpha$ narrow component emission line map (see
Fig\,~ \ref{fig:Hamaps}). North is up and east is to the left.  H$\alpha$
contours are shown for the range 24.8 -- 899.9, in steps of 119, in surface
brightness units of $\times10^{-15}$ ergs cm$^{-2}$ s$^{-1}$ arcsec$^{-2}$.}
\label{fig:HSToverlay}
\end{center}
\end{figure*}

The current study focuses on Markarian 996, a BCD whose properties are far from
ordinary. \citet{Thuan:1996} (TIL96 hereafter) presented a comprehensive {\it
HST} study of Mrk~996 by analysing FOS UV and optical spectra and WFPC2 $V$-
and $I$-band imaging. Their {\it HST} $V$-band image, shown in Figure
\ref{fig:HSToverlay}, revealed that the majority of the star formation in
Mrk~996 occurs within a bright, compact nuclear \hii\ region of $\sim$3 arcsec
angular radius ($\sim$315 pc for their adopted distance of 21.6 Mpc (TIL96))
surrounded by an extended elliptical low surface brightness component of
$\sim$40.6 arcsec diameter. However, the remarkable nature of the galaxy was
exposed by the small-aperture ($\sim$1 arcsec) FOS spectra obtained from the
nucleus of Mrk~996. Line widths increase with the degree of ionisation; lines
from singly ionised species being narrow and typical of \hii\ regions whereas
lines from ions such as O$^{2+}$ and Ne$^{2+}$ are very broad, with FWHM widths
up to $\sim$900 km s$^{-1}$ (TIL96). Un-physically high [O\three] electron
temperatures are derived for the nucleus unless an electron density (\eld) of
at least $\sim10^6$ cm$^{-3}$ is adopted, several orders of magnitude higher
than normal \hii\ region electron densities \citep[][]{Izotov:1994}. In order
to account for the observed line intensities, TIL96 applied a two-zone
density-bounded \hii\ region model, with an inner core density of $\sim10^6$
cm$^{-3}$ and an outer zone with a density of $\sim450$ cm$^{-3}$. They
suggested that this large density gradient is caused by a mass outflow driven
by the large population of Wolf-Rayet stars found to be present in the galaxy.

Elemental abundances were also derived by TIL96, who estimated a metallicity of
0.22\,\Zsol, based on the abundance of oxygen.  As a result of their study,
Mrk~996 was placed in a small group of eight BCDs known to display a
significant overabundance of nitrogen \citep[][]{Pustilnik:2004}, with derived
N/O ratios that are 0.3--1.4 dex higher than for other BCDs of similar oxygen
abundances. The enhanced nitrogen within this group has been suggested by
\citet{Pustilnik:2004} to be connected with merger events, in particular with a
short phase of the consequent starburst, when many WR stars contribute to the
enrichment of the interstellar medium (ISM).   This was certainly the case for NGC~5253 (thought to be interacting with its companion galaxy NCG~5236 \citep{Moorwood:1982}), where spatially resolved N/H and N/O enhancements were observed by \citet{Walsh:1987} to correlate strongly with WR star signatures.  In support of this, a correlation was also found between spatial enhancements of He/H and N/O, attributed to helium enrichment of the ISM by WR stars \citep{Walsh:1989}.

The study of TIL96 showed that Mrk~996 has spatially varying physical
properties. Spatially resolved kinematic and chemical abundance information
across the galaxy is therefore crucial for understanding the nature of this
system; such information can be readily obtained with integral field
spectroscopy (IFS). In this paper we present high resolution optical IFS
observations obtained with the VIMOS integral field unit (IFU) spectrograph on
the 8.2m Very Large Telescope UT3/Melipal. The data afford us a new
spatiokinematic `3-D' view of Mrk~996. The spatial and spectral
resolution achieved allows us to undertake a full multi-velocity-component
analysis of this system, ultimately providing a more complete picture of its
diverse ionised ISM.

We adopt a distance of 22.3 Mpc for Mrk~996, at a redshifted velocity of 1642
km s$^{-1}$ (corresponding to $z$ $=$ 0.00544, this work), using a Hubble
constant H$_{\rm o}$ $=$ 73.5 km s$^{-1}$~Mpc$^{-1}$
\citep[][]{Bernardis:2008}.

\section[]{VIMOS IFU Observations and Data Reduction}
\subsection{Observations}

Two data sets were obtained with the Visible
Multi-Object Spectrograph (VIMOS) IFU at the 8.2m VLT at ESO's Paranal
Observatory in Chile. The VIMOS IFU consists of 6400 (80$\times$80) fibres
coupled to microlenses and operates with four CCD detectors yielding four
spectral and spatial quadrant samplings \citep{zanichelli:2005}.  The spatial
sampling is contiguous, with the dead space between fibres below 10 per cent of
the inter-fibre distance.  The data sets were taken with the high
resolution and high magnification (i.e. a partially masked IFU head containing
only 40$\times$40 fibres) settings of the IFU, resulting in a field-of-view
(FoV) of 13$^{\prime\prime}$ $\times$ 13$^{\prime\prime}$ covered by 1600
spatial pixels (spaxels) at a spatial sampling of 0.33$^{\prime\prime}$ per
spaxel. Two different grisms were used, high resolution blue (HRblue, $\sim$0.51\,\AA/pixel), covering 4150--6200\,\AA\ and high resolution
orange (HRorange, $\sim$0.60\,\AA/pixel) covering 5250--7400\,\AA;
spectra from both grisms are illustrated in Fig\,~\ref{fig:fullSpec}.  The instrumental width delivered by the IFU HRblue and HRorange grisms was
measured using arc lamp exposures. Gaussian profiles were fitted to a number of
isolated lines within each extracted arc spectrum and were found to have FWHMs
of 2.3 $\pm $0.1\,\AA\ (113.5 $\pm$ 4.9\,km s$^{-1}$) and 1.5 $\pm
$0.1\,\AA\ (72.3 $\pm$ 4.8 km s$^{-1}$) for the HRblue and HRorange grisms,
respectively.

The observing log can be found in Table~\ref{tab:observations}.  Four exposures
were taken per grism.  The third exposure within each set was
dithered by $+$0.34$^{\prime\prime}$ in RA and by $+$0.58$^{\prime\prime}$ in
Dec (corresponding to a 2 spaxel offset in the $X$-direction) in order to
remove any dead fibres when averaging exposures. All observations were taken at
a position angle of $+$30$^\circ$.

\subsection{Data Reduction}
\label{sec:reduction}

The data reduction was carried out using the GUI-based pipeline software {\sc
gasgano}{\footnote{http://www.eso.org/sci/data-processing/software/gasgano}
that allows the user to organise calibration files and run pipeline tasks.
Three main tasks were used: (1) {\it vmbias} created the master bias frame; (2)
{\it vmifucalib} determined the spectral extraction mask, wavelength
calibration and the relative fibre transmission correction; (3) {\it
vmifustandard} created flux response curves from a summed spectrophotometric
standard star spectrum. Finally, the products of these tasks were fed into {\it
vmifuscience} which extracted the bias-subtracted, wavelength- and
flux-calibrated, and relative fibre transmission-corrected science spectra.
Each VIMOS CCD quadrant is an independent spectrograph, therefore the IFU data
processing was performed separately on each quadrant, creating four fully
calibrated 3-D arrays per science exposure. The flux calibration was performed
by multiplying the 2-D spectral frame for each quadrant by its relative
response curve, derived from standard star observations in the same filter,
grism and quadrant. The final data cube is produced by median-combining each
exposure in a jitter sequence (i.e. re-aligning the dithered exposure with the
other exposures of identical pointing), after all the individual data quadrants
for each exposure had been reduced. Reconstructing the cube spatially utilises
the IFU table that lists the one-to-one correspondence between fibre positions
on the IFU head and stacked spectra on the CCD \citep{Bastian:2006}.  A schematic representation of
the data reduction processes can be found in \citet{zanichelli:2005}.

\begin{table*}
\begin{center}
\begin{small}
\caption{VIMOS IFU observing log}
\begin{tabular}{|cccccc|}
\hline
Observation ID & Date &Grism    &Exp. time (s)& Airmass range & FWHM Seeing (arcsec)\\
\hline
250520 & 13/10/2006 & HR blue & 4 ${\times}$ 402 & 1.135 -- 1.097 & 0.72 \\
250516 & 19/10/2006 & HR orange & 4 ${\times}$ 372 & 1.057 -- 1.053 & 1.15\\
\hline \label{tab:observations}
\end{tabular}
\end{small}
\end{center}
\end{table*}

Unlike other IFU designs (such as VLT FLAMES/Argus), there are no
`sky-dedicated' fibres in VIMOS IFU mode. Sky subtraction was performed by
locating a background region within each quadrant, summing the spectra over
each of the spaxels in the reconstructed cube and subtracting the median sky
spectrum of the region from its corresponding quadrant.  Median combining was
needed to ensure that any residual contamination from faint objects was
removed. It should be noted that VIMOS IFU images are obtained as projections
on the IFU CCD, i.e. they are mirror images to what is seen on the sky. Thus,
all emission maps presented in this paper have been reflected about the
$Y$-axis to obtain the correct orientation.

\begin{figure*}
\begin{center}
\includegraphics[scale=0.73, angle=90]{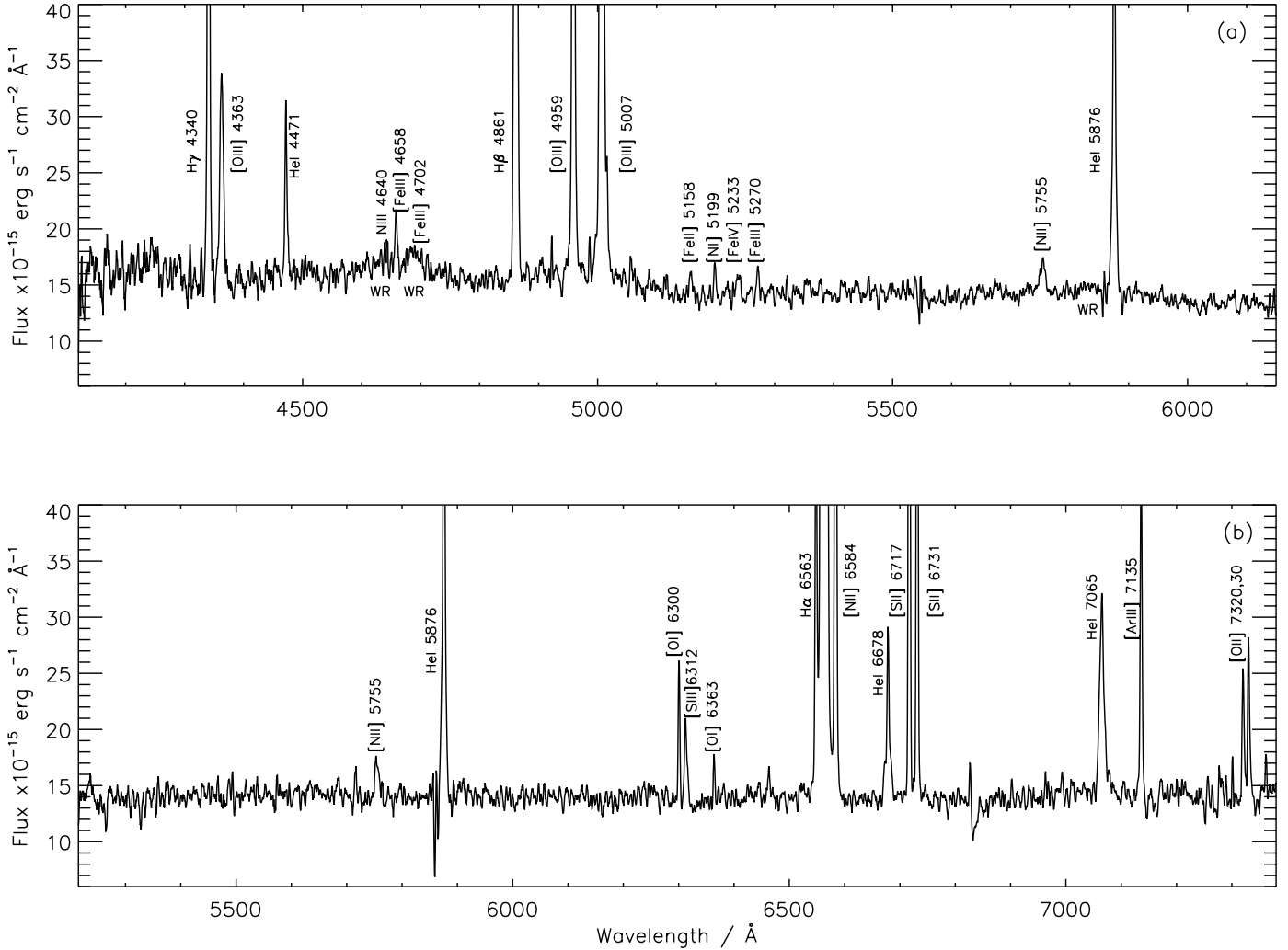}
\caption{VIMOS IFU summed spectra of Mrk~996 after smoothing with a 5 pixel
boxcar: (a) high resolution blue spectrum; (b) high resolution orange spectrum.
Spectra are integrated over an area of 5.3$\times$6.3 arcsec$^2$
(575$\times$685 pc$^2$) and correspond to exposure times of 4$\times$402 s and
4$\times$372 s for the HR blue and HR orange grisms, respectively.}
\label{fig:fullSpec}
\end{center}
\end{figure*}

\subsection{Emission Line Profile Fitting}

The resolving power and signal-to-noise (S/N) of the data are high enough to
allow the resolution of multiple components in the majority of the emission
lines seen in the 1600 spectra across the FoV.  In order to fit spectra of this
quantity and complexity, we utilised an automated fitting procedure called PAN
\citep[Peak ANalysis; ][]{dimeo:2005}.  This IDL-based, general-purpose
curve-fitting package was adapted by \citet{Westmoquette:2007a} for use with
FITS format data by allowing multiple spectra to be read simultaneously in an
array format. The user can interactively specify initial parameters of a
spectral line fit (continuum level, line peak flux, centroid and width) and
allow PAN to sequentially process each spectrum, fitting Gaussian profiles
accordingly.  The output consists of the fit parameters for the continuum and
each spectral line's profile and the $\chi^2$ value for the fit.

It was found that in the spectra of Mrk~996 most high S/N ratio emission line
profiles consist of a strong, narrow central Gaussian component (hereafter C1),
superimposed on a very broad component containing a substantial proportion of the flux
(hereafter C2). In addition to this, in the core of the galaxy the C2 component
of the Balmer line profiles splits into two broad Gaussians of variable
strength and velocity (hereafter C3 and C4, shifted towards the blue and red of
the centroid of C1, respectively). The splitting of C2 into C3 and C4 and the
region in which this occurs is discussed in Section~\ref{sec:HaMaps}.
Fig\,~\ref{fig:fit_comps} shows examples of fits to the H$\alpha$ emission line
from different regions of Mrk~996 in order to illustrate single, double and triple
Gaussian fits, along with the corresponding fit residuals. In the same figure a
double Gaussian fit to the [O\three] $\lambda$5007 profile is also shown
corresponding to the same central spaxel for which H$\alpha$ requires a triple
Gaussian fit, thus illustrating the difference between the H$\alpha$ and [O\three] line profiles.
Where appropriate, single or multiple Gaussians were fitted to the emission
line profiles, restricting the minimum FWHM to be the instrumental width.
Suitable wavelength limits were defined for each emission line and continuum
level fit. Further constraints were applied when fitting the [S\two] doublet:
the wavelength difference between the two lines was taken to be equal to the redshifted
laboratory value when fitting the velocity component, and their FWHMs were set
equal to one another.

In order to rigorously determine the optimum number of Gaussians required to
fit each observed profile the statistical F-test was used. The F-distribution
function allows one to calculate the significance of a variance ($\chi^2$)
increase that is associated with a given confidence level, for a given number
of degrees of freedom.  We note that even though many low S/N ratio lines may
in reality be composed of multiple emission components, it is often not
statistically significant to fit anything more than a single Gaussian. This is
why the robust F-test method was chosen for objectively determining the optimum
number of components to fit to each line. Resultant optimised fits for various
line species are as follows:

{\it Single narrow} Gaussian (i.e. C1 component only) -- [Fe~{\sc iii}]
$\lambda\lambda$4658, 4702, 4881, 4986, [O\one] $\lambda\lambda$6300, 6364,
[N\two] $\lambda$6584, [S\two] $\lambda\lambda$6717, 6731, and [O\two]
$\lambda\lambda$7320, 7330.

{\it Single broad} Gaussian (i.e. C2 component only) -- [N\two] $\lambda$5755,
[O\three] $\lambda$4363.

{\it Double narrow/broad} Gaussian (C1 and C2) -- all Balmer lines (excluding
the central spaxels), \hei\ $\lambda\lambda$4471, 5876, 6678, 7065, [O\three]
$\lambda\lambda$4959, 5007, [S\three] $\lambda$6312, and [Ar\three]
$\lambda$7136.

{\it Triple narrow/broad} Gaussian fit (C1, C3, and C4) -- all Balmer lines
(central spaxels only).

The errors reported by PAN during fitting underestimate the true uncertainties.
We thus follow an error estimation procedure which involves the visual
re-inspection of the line profile plus fit after knowing which solution was
selected by our tests, and taking into account the S/N ratio of the spectrum.
The errors on the fits are minimised across the IFU's aperture varying S/N
ratio during the F-testing procedure by selecting the number of components
which best-fit the profile within that particular spaxel. It is further noted
that the uncertainties associated with each individual line component are
coupled to those of the other components within a given line profile. By
comparing PAN fits to those performed by other fitting techniques (e.g. {\sc
iraf}'s {\sc splot} task) on line profiles with an established configuration
(i.e. after performing the F-test) we find that uncertainties of $\sim$5--10
and $\sim$15--20 per cent are associated with the C1 and C2--4 fits,
respectively. A listing of measured flux and FWHM uncertainties appears in
Table \ref{tab:fluxvals}, where errors are quoted for individual component fits
to emission lines detected on integrated spectra across the inner and outer
regions of Mrk~996.

\begin{figure*}
\begin{center}
\includegraphics[scale=0.75]{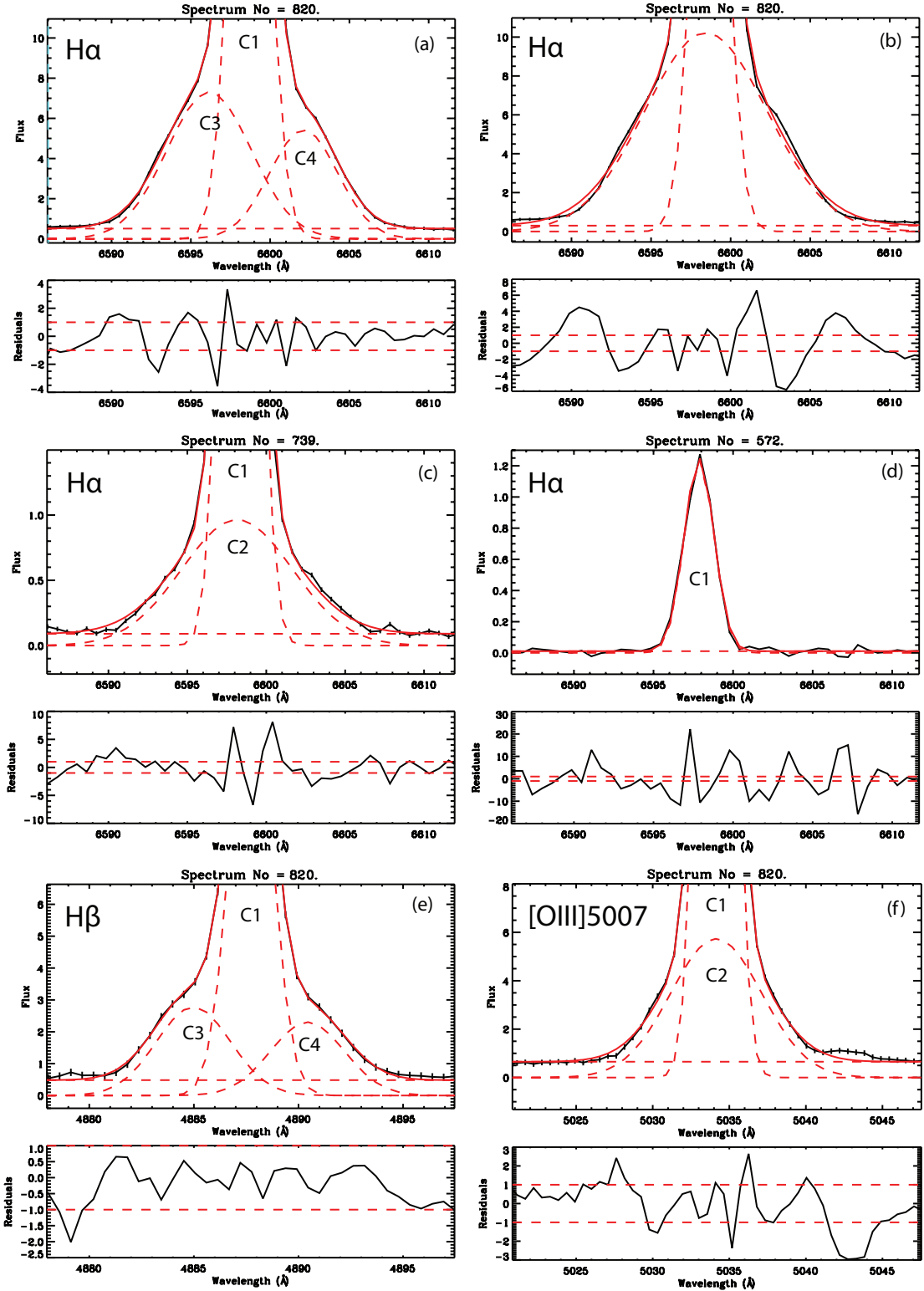}

\caption{Example Mrk~996 emission line profiles and Gaussian fits: (a) a
typical spaxel from the core region with a triple Gaussian fit to the H$\alpha$
line;  (b) the same spaxel spectrum shown in (a) but with a double Gaussian fit
[note the larger residuals of the fit as compared to (a)]; (c) H$\alpha$
emission line from a spaxel in the outer region which required a double
Gaussian fit; (d) H$\alpha$ emission line from the edge of the outer region
where a single component fit is sufficient;  (e) the H$\beta$ emission line for
the same core spaxel as that in (a), displaying the same triple component
structure (H$\gamma$ also shows an identical structure in the core spaxels);
(f) the [O\three] $\lambda$5007 emission line for the same core spaxel as that
of (a) and (b), illustrating an optimal double Gaussian fit (rather than the
triple components required for the Balmer lines). The flux units are
$\times10^{-15}$erg s$^{-1}$ cm$^{-2}$ {\AA}$^{-1}$.  The fit residuals are
plotted under each panel (in units of $\sigma$) with dashed guidelines at
$\pm$1$\sigma$.  The wavelengths are not corrected for redshift.}
\label{fig:fit_comps}
\end{center}
\end{figure*}

\subsection{Cube alignment and correction for differential atmospheric refraction}

Observations of an object's spectrum through the Earth's atmosphere are subject
to refraction as a function of wavelength, known as differential atmospheric
refraction, DAR. The direction of DAR is along the parallactic angle at which
the observation is made. Each reduced data cube was corrected for this effect
using the algorithm outlined by \citet{Walsh:1990}; this procedure calculates
fractional pixel shifts for each monochromatic slice of the cube relative to a
fiducial wavelength (i.e. a strong emission line), shifts each slice with
respect to the orientation of the slit on the sky and the parallactic angle and
re-combines the DAR-corrected data cube.  Having data cubes with different
wavelength ranges would ideally involve a choice of fiducial wavelength that is
common to both, e.g. \hei\ $\lambda$5876 for these VIMOS data. However, the S/N
ratio of this line is not consistently high across the IFU aperture, and pixel
shifts for each cube relative to this line were found to be unreliable.
Instead, we chose to correct each data cube using its strongest emission line:
H$\beta$ for the HRblue cube and H$\alpha$ for the HRorange cube.  The two
cubes were then spatially matched by aligning the H$\alpha$ and H$\beta$
emission contours using a 2-D Gaussian fit.  This involved a maximum shift of
$-$0.73 spaxels in $X$ and $+$0.82 spaxels in $Y$. Two effects can introduce
minor errors when aligning monochromatic \hi\ images: (i) an extinction
gradient across the observed region disproportionately affecting the two lines,
and (ii) the presence of a temperature gradient, since the \hi\ lines have a
very slightly different temperature dependence. However, both gradients would
need to be significant for substantial errors to be introduced; the sub-spaxel
shifts that were determined show that this is not the case here.

\begin{table*}
\caption{Emission line measurements for summed spectra of the inner core
(1.7$''\times$2.3$''$) and an `outer annulus' region whose extent is
5.3$''\times$6.3$''$ but from which the inner core spectrum has been subtracted
(see Section~\ref{sec:broads}). Observed fluxes and dereddened intensities are
given for the main two velocity components and are relative to H$\beta$ $=$ 100
(the corresponding velocity component of H$\beta$). The employed logarithmic
extinction constants, $c$(H$\beta$), are quoted (see
Section~\ref{sec:reddening}); integrated H$\beta$ fluxes for the two main
velocity components, in units of $\times$10$^{15}$ erg cm$^{-2}$ s$^{-1}$, are
also listed.}

\begin{center}
\begin{small}
\begin{tabular}{lc ccc ccc}
\noalign{\vskip3pt}\noalign{\hrule}\noalign{\vskip3pt}
 &  &\multicolumn{3}{c}{Core Region} & \multicolumn{3}{c}{Outer Region}\\
 Line ID & Component & FWHM (km s$^{-1}$) & $F$($\lambda$) & $I$($\lambda$) & FWHM (km s$^{-1}$) & $F$($\lambda$) & $I$($\lambda$)\\
\noalign{\vskip3pt}\noalign{\hrule}\noalign{\vskip3pt}
 4340        H$\gamma$& C1 &  145.1$\pm$ 2.1 &    44.9$\pm$    0.5 &     46.3 $\pm$     0.4  &       158.2$\pm$       6.2 &  54.4$\pm$    8.7 &     63.2 $\pm$     2.2 \\
 4340        H$\gamma$& C2 &  442.2$\pm$ 15.9 &    42.6$\pm$    1.3 &     48.2 $\pm$     1.4  &       -- &       -- &   --\\
 4363    [O\three]& C2 &      476.5$\pm$ 14.4 &    20.2$\pm$    0.5 &     32.9 $\pm$     0.8  &       -- &       -- &   --\\
 4471        He\one& C1 &     155.6$\pm$ 14.8 &     3.8$\pm$    0.6 &      3.9 $\pm$     0.6  &       260.2$\pm$       91.2 &   9.7$\pm$    2.9 &     10.8 $\pm$     2.8 \\
 4471        He\one& C2 &     540.6$\pm$ 61.0 &     9.8$\pm$    1.0 &     10.7 $\pm$     1.1  &       -- &       -- &   --\\
 4658        [Fe\three]& C1 & 277.5$\pm$ 27.7 &     3.3$\pm$    0.2 &      3.3 $\pm$     0.5  &       -- &       -- &   --\\
 4702    [Fe\three]& C1 &     274.3$\pm$ 53.0 &     0.7$\pm$    0.4 &      0.7 $\pm$     0.4  &       -- &       -- &   --\\
 4861        H$\beta$& C1 &   130.8$\pm$ 0.6 &   100.0$\pm$    1.0 &    100.0 $\pm$     0.7 &       133.9$\pm$      0.6 & 100.0$\pm$   22.0 &    100.0 $\pm$    15.6 \\
 4861    H$\beta$& C2 &       414.7$\pm$ 4.3 &   100.0$\pm$    1.5 &    100.0 $\pm$     1.0  &       308.5$\pm$       6.2 & 100.0$\pm$   21.5 &    100.1 $\pm$    15.2 \\
 4881    [Fe\three]& C1 &     313.4$\pm$ 94.6 &   1.0$\pm$    0.4 &      1.0 $\pm$     0.4    &       -- &       -- &   --\\
 4959        [O\three]& C1 &  131.2$\pm$ 0.6 &   132.2$\pm$    1.3 &    131.7 $\pm$     0.9 &       137.9$\pm$1.75 &113.1$\pm$  2.3 & 109.9$\pm$  2.2 \\
 4959        [O\three]& C2 &  392.0$\pm$ 10.3 &    49.3$\pm$    1.4 &     48.2 $\pm$     1.3  &       332.7$\pm$ 3.6 &   58.32 $\pm$ 14.8 & 55.4 $\pm$ 14.08 \\
 4986        [Fe\three]& C1 & 210.5$\pm$ 54.8 &     1.5$\pm$    0.6 &      1.5 $\pm$     0.6  &       -- &       -- &   --\\
 5007        [O\three] & C1 & 132.4$\pm$ 0.6 &   399.6$\pm$    4.2 &    397.1 $\pm$     3.2 &       130.6$\pm$      0.6 &  347.7$\pm$   54.3 &    333.4 $\pm$     2.5  \\
 5007        [O\three]& C2 &  415.2$\pm$ 13.8 &   148.8$\pm$    4.7 &    143.8 $\pm$     4.3    &   404.4$\pm$       38.9 & 133.9$\pm$   24.1 &    124.3 $\pm$    11.9  \\
 5755        [N\two]& C2 &    427.9$\pm$ 74.5 &     3.3$\pm$    0.4 &      4.0 $\pm$     0.5  &       -- &       -- &   --\\
 5875        He\one& C1 &     121.5$\pm$ 2.5 &    11.3$\pm$    0.4 &     10.7 $\pm$     0.4  &       147.0$\pm$       8.7 &  19.6$\pm$    3.2 &     15.2 $\pm$     0.7 \\
 5875        He\one& C2 &     425.8$\pm$ 8.2 &    33.4$\pm$    0.7 &     27.2 $\pm$     0.5  &       -- &       -- &    --\\
 6300    [O\one]& C1 &        128.0$\pm$ 13.8 &    4.6$\pm$    0.4 &      4.3 $\pm$     0.4   &       141.9$\pm$       22.9 &   7.6$\pm$    1.5 &      5.4 $\pm$     0.7 \\
 6312        [S\three]& C1 & 107.4$\pm$ 13.8 &     1.2$\pm$    0.2 &      1.2 $\pm$     0.2    &       133.0$\pm$       31.8 &   3.8$\pm$    0.8 &      2.7 $\pm$     0.4 \\
 6312        [S\three]& C2 &403.0$\pm$ 29.0 &     5.8$\pm$    0.4 &      4.4 $\pm$     0.3  &       -- &       -- & --\\
 6363        [O\one] & C1 &   153.2$\pm$ 20.3 &     1.7$\pm$    0.2 &      1.6 $\pm$     0.2  &       147.5$\pm$       38.2 &   4.0$\pm$    1.1 &      2.9 $\pm$     0.6 \\
 6563        H$\alpha$& C1 & 116.5$\pm$ 0.9 &   306.9$\pm$    4.4 &    285.6 $\pm$     3.6 &       118.8$\pm$      0.4 & 418.9$\pm$   65.3 &    287.6 $\pm$     1.6 \\
 6563        H$\alpha$& C2 &  420.5$\pm$ 7.3 &   400.5$\pm$    7.5 &    294.0 $\pm$     4.6  &       407.7$\pm$       9.6 & 555.5$\pm$   85.3 &    284.2 $\pm$     0.8 \\
 6584    [N\two]& C1 &        126.2$\pm$ 4.1 &   38.0$\pm$    1.1 &     35.3 $\pm$     1.0   &       123.4$\pm$       2.2 &   45.4$\pm$    7.1 &     31.1 $\pm$     0.4 \\
 6678        He\one& C1 &     112.7$\pm$ 4.9 &     3.5$\pm$    0.2 &      3.2 $\pm$     0.2  &       124.8$\pm$       14.8 &    6.8$\pm$    1.3 &      4.6 $\pm$     0.5 \\
 6678    He\one& C2 &         550.3$\pm$ 28.8 &     9.2$\pm$    0.4 &      6.6 $\pm$     0.3  &       -- &       -- &   --\\
 6717        [S\two]& C1 &    118.8$\pm$ 1.8 &    30.9$\pm$    0.5 &     28.6 $\pm$     0.5  &       117.4$\pm$       2.2 &  31.6$\pm$    5.0 &     21.2 $\pm$     0.4 \\
 6731        [S\two]& C1 &    118.5$\pm$ 1.8 &    24.6$\pm$    0.5 &     22.8 $\pm$     0.4  &       117.2$\pm$       2.2 &   22.4$\pm$    3.5 &     14.9 $\pm$     0.4 \\
 7065        He\one& C1 &     76.00$\pm$ 65.4 &     1.1$\pm$    0.9 &      1.0 $\pm$     0.8  &       217.8$\pm$       56.1 &    7.0$\pm$    1.9 &      4.4 $\pm$     1.0 \\
 7065        He\one& C2 &     425.4$\pm$ 58.2 &     24.4$\pm$    2.8 &     16.8 $\pm$     1.9 &       -- &       -- &   --\\
 7136        [Ar\three]& C1 & 127.3$\pm$ 17.2 &    10.3$\pm$    2.3 &      9.4 $\pm$     2.1  &       127.8$\pm$       11.4 &  13.5$\pm$    2.3 &      8.5 $\pm$     0.7 \\
 7136        [Ar\three]& C2 & 441.0$\pm$ 213.9 &      8.2$\pm$    3.5 &      5.6 $\pm$     2.4 &       -- &       -- &  --\\
 7320        [O\two]& C1 &    176.6$\pm$ 11.9 &     6.1$\pm$    0.3 &      5.6 $\pm$     0.3  &       219.2$\pm$       62.3 &   9.2$\pm$    2.3 &      5.6 $\pm$     1.1 \\
 7330        [O\two]& C1 &    165.3$\pm$ 11.9 &     5.8$\pm$    0.3 &      5.3 $\pm$     0.3  &       111.3$\pm$       21.3 &   10.8$\pm$    2.2 &      6.6 $\pm$     0.8 \\
    \noalign{\vskip3pt}
$c$(H$\beta$)       & C1  &\multicolumn{3}{c}{0.10$\pm$0.02} & \multicolumn{3}{c}{0.51$\pm$0.08}\\
$c$(H$\beta$)       & C2  &\multicolumn{3}{c}{0.42$\pm$0.04} & \multicolumn{3}{c}{0.91$\pm$0.12}\\
$F$(H$\beta$) & C1  & \multicolumn{3}{c}{626.30$\pm$4.31} & \multicolumn{3}{c}{180.33$\pm$28.13} \\
$F$(H$\beta$) & C2  & \multicolumn{3}{c}{431.70$\pm$4.46} & \multicolumn{3}{c}{36.67$\pm$5.57} \\
\noalign{\vskip3pt}\noalign{\hrule}\noalign{\vskip3pt}
\end{tabular}
\end{small}
\end{center}

\label{tab:fluxvals}
\end{table*}

\section{Mapping of line fluxes and kinematics}

\subsection{Line fluxes and reddening correction}
\label{sec:reddening}

Full HRblue and HRorange spectra are shown in Fig\,~\ref{fig:fullSpec}, with
identified emission lines labelled.  Table~\ref{tab:fluxvals} lists the
measured FWHM and observed and de-reddened fluxes for the main velocity
components of the detected emission lines.  The fluxes are from spectra summed
over the core region and the outer annulus region, as defined in
Section~\ref{sec:broads}, and are quoted relative to the flux of the
corresponding H$\beta$ component. They were corrected for reddening using the
Galactic reddening law of \citet{Howarth:1983} using $c$(H$\beta$) values
derived from the H$\alpha$/H$\beta$ and H$\gamma$/H$\beta$ line ratios of their
corresponding components, weighted in a 3:1 ratio, respectively, in conjunction
with the theoretical Case B ratios from \citet{Hummer:1987}. Following the same
method, an average $c$(H$\beta$) map was also derived using ratioed
H$\alpha$/H$\beta$ and H$\gamma$/H$\beta$ emission line maps.  The
$c$(H$\beta$) map was used for correcting emission line maps for reddening in
order to then create abundance maps. In the direction to Mrk~996 a foreground
Milky Way reddening of $E$($B-V$) $=$ 0.04, corresponding to $c$(H$\beta$) $=$
0.06, is indicated by the extinction maps of \citet{Schlegel:1998}. Values of
$c$(H$\beta$) of 0.10$\pm$0.02 and 0.51$\pm$0.08 are applicable to the narrow
component emission in the inner and outer regions of Mrk~996, respectively,
while values of 0.42$\pm$0.04 and 0.91$\pm$0.12 are applicable to the broad
component emission for the inner and outer regions, respectively.

\subsection{The Balmer line properties}
\label{sec:HaMaps}

\begin{figure*}
\begin{center}
\includegraphics[scale=0.8]{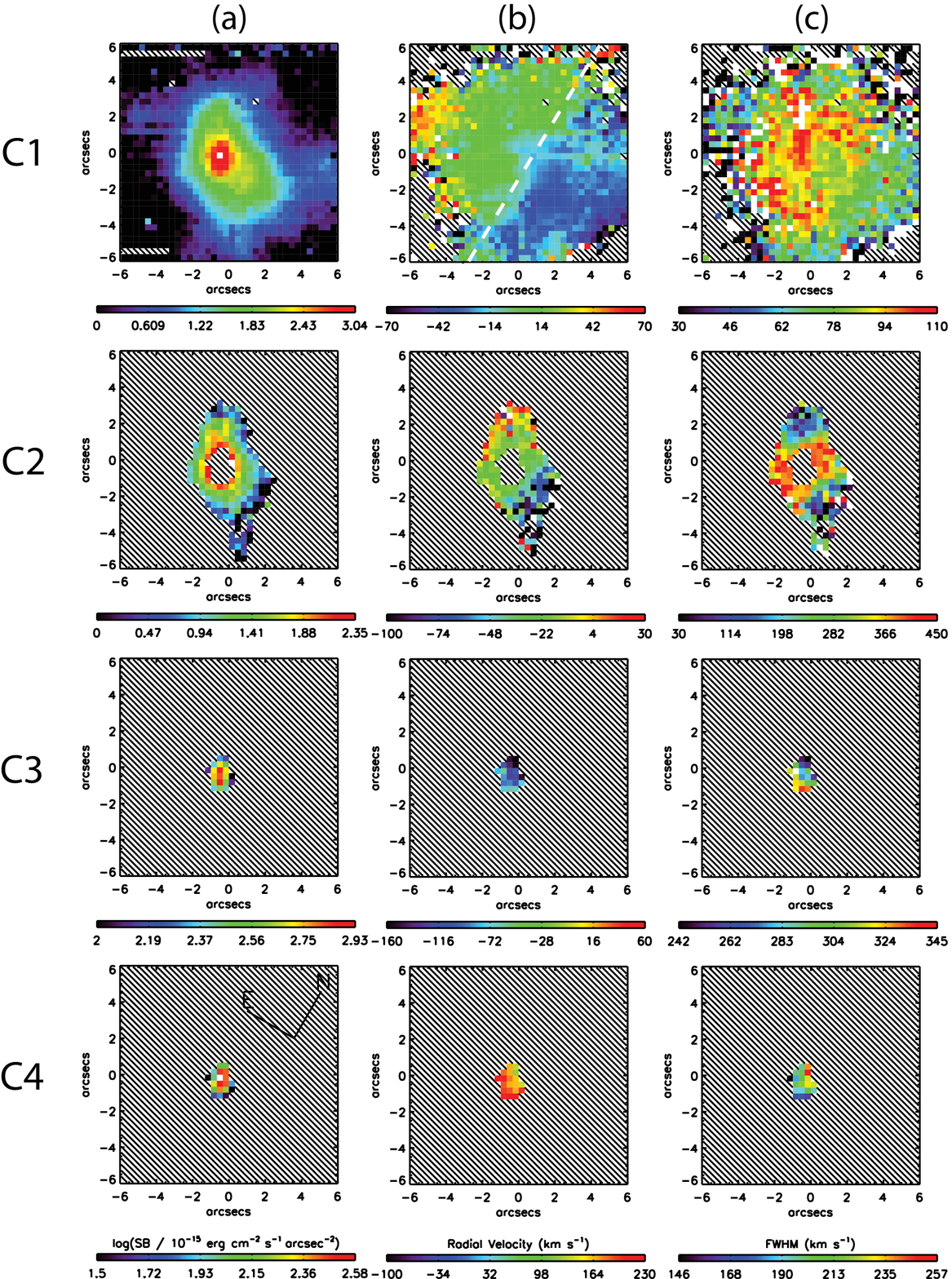}

\caption{Maps of Mrk~996 in the H$\alpha$ velocity components C1--4; (a)
logarithmic surface brightness; (b) radial velocity (relative to the
heliocentric systemic velocity of $+$1642 km s$^{-1}$), where the white dotted line overlaid on the C1 velocity map represents a cut along the rotational axis, as used for the position-velocity diagrams shown in Fig~\ref{fig:velcuts}; (c) FWHM corrected for
the instrumental PSF. See text for details.} \label{fig:Hamaps}
\end{center}
\end{figure*}

All the Balmer lines detected from Mrk~996 (H$\alpha$, H$\beta$ and H$\gamma$)
show a common multi-component velocity structure which in the innermost spaxels
is exclusive to the Balmer lines, i.e. it is not seen in the forbidden lines.
Here we present a discussion of the highest S/N ratio \hi\ line, H$\alpha$.
Maps of the individual H$\alpha$ velocity components (flux, radial velocity and
FWHM) are shown in Fig\,~\ref{fig:Hamaps}. As mentioned previously, up to three
components are resolved in the H$\alpha$ line profile in the nuclear region of
Mrk~996:

(i) A central, narrow Gaussian (C1) is detected throughout the galaxy (see the
flux map in Fig\,~\ref{fig:Hamaps}a), peaking at $\alpha = 01^{\rm h}\, 27^{\rm
m}\, 35\fsec5\pm0.1$, $\delta = -06^\circ\,19'\,36\farcs2\pm0.2$ (J2000), with
a surface brightness of 10.1$\times$10$^{-12}$ erg cm$^{-2}$ s$^{-1}$
arcsec$^{-2}$ in the central region, and gradually decreasing over a region
with a major axis radius of 3$''$.84 (403 pc) and a minor axis of 2$''$.39 (251
pc); beyond this the H$\alpha$ emission traces a very low surface brightness
area (which TIL96 fitted with an exponential disk on their {\it HST} images).
The galaxy does not show a smooth elliptical shape when viewed in H$\alpha$.
Fig\,~\ref{fig:HSToverlay} shows the flux contours of the narrow H$\alpha$
component (C1) overlaid on the {\it HST} F569W ({\it V}-band) image. The H$\alpha$ emitting
region imaged by VIMOS is mostly confined within the inner part of the F569W
image but the major axes of the two are not aligned. A dust extinction feature
is present in the north-west section of the inner star forming region that
distorts the F569W flux isophotes (TIL96). The dust feature also distorts the
outermost H$\alpha$ contours of the VIMOS data.

(ii) In the very core of Mrk~996, over an area of $\sim$1$''$.5 in radius (162
pc), the narrow C1 profiles are accompanied by broad underlying components that
are best fit with two Gaussians (C3 and C4) to the blue and red of C1,
respectively (Fig\,~\ref{fig:Hamaps}a). The significance of these features is
discussed below.

(iii) Beyond a radius of $\sim$1$''$.5 from the core, C3 and C4 blend into a
single broad Gaussian (C2) (Fig\,~\ref{fig:Hamaps}a). The region over which
this component is seen is $\sim$2--4 arcsec in radius (216--432 pc); its outer
boundary matches well that of the brightest portion of the C1 flux map.


\begin{figure*}
\begin{center}
\includegraphics[height=7.5cm,width=17cm]{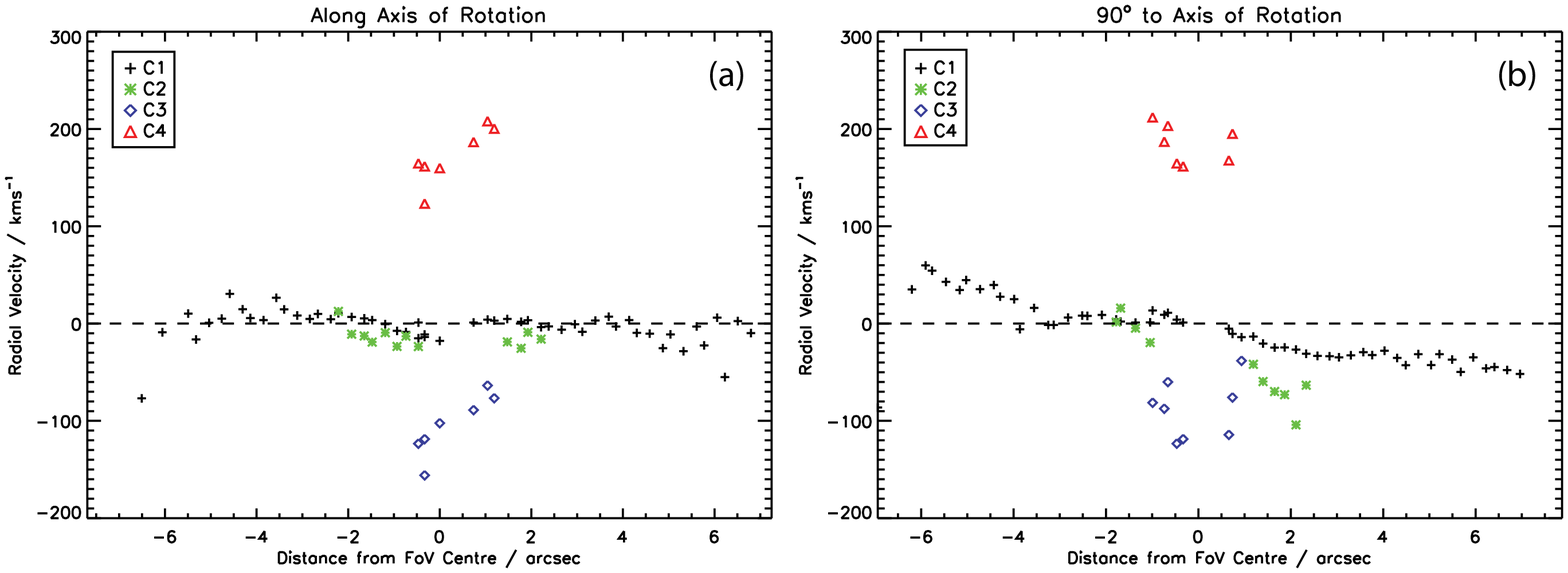}

\caption{Position velocity diagrams for H$\alpha$: (a) along the axis of
rotation, defined as 5$^\circ$ east of north, as indicated by component C1 in
Fig\,~\ref{fig:Hamaps}b; (b) along the direction 90$^\circ$ to the rotation
axis. Both diagrams are relative to a systemic velocity of $+$1642 km s$^{-1}$.
Typical radial velocity errors are $<$3, 8--20, 12--15, and 7--11 per cent for
C1, C2, C3 and C4, respectively, with a spatial error of 0.5~spaxels
($\pm$0.17$''$).} \label{fig:velcuts}
\end{center}
\end{figure*}

The radial velocity map of component C1 (Fig\,~\ref{fig:Hamaps}b) shows a
strong velocity gradient from west to east, with a definable axis of rotation
at PA $= 5^{\circ}$ (marked by the dashed line across the C1 velocity map).
Neither the rotation axis as revealed by VIMOS nor the axis normal to it are
aligned with the apparent, i.e. photometric, major axis of the galaxy which lies at PA
$\sim23^{\circ}$ (TIL96). A velocity gradient in approximately the same direction is
observed in the broad component C2, although in this case the negative velocity
component is not exactly balanced by a positive counterpart.
Fig\,~\ref{fig:velcuts} shows position-velocity (P-V) diagrams along the
adopted axis of rotation (Fig\,~\ref{fig:velcuts}a) and also perpendicular to
this axis (Fig\,~\ref{fig:velcuts}b). A new estimation of the heliocentric
systemic velocity of Mrk~996 can be made using Fig\,~\ref{fig:velcuts}a; a
radial velocity of 1642 $\pm$ 10 km s$^{-1}$ is needed to normalize the
distribution of C1 velocities along the axis of rotation to zero (cf. the 1622
$\pm$ 10\,km s$^{-1}$ estimate of TIL96). The narrow line component C1 shows a
velocity gradient along the direction normal to the rotation axis, with radial
velocities ranging from $\sim-$50\,km s$^{-1}$ to $+$50\,km s$^{-1}$, and the
P-V diagram of component C1 shows a linear velocity gradient indicating
solid-body rotation (Fig\,~\ref{fig:velcuts}b). This type of structure is not
uncommon for BCDs observed in the \hi\ 21 cm line and has been used by
\citet{VanZee:2001} to constrain evolutionary
scenarios.

Assuming a Keplerian velocity profile we have converted each radial velocity
data point of component C1 (Fig\,~\ref{fig:velcuts}b) into a mass within radius
$r$, as shown in Fig\,~\ref{fig:dynMass}. This plot shows dynamical mass
growing at similar rates on either side of the centre, rising to a total mass
of $\sim5\pm$1$\times$10$^8$M$_{\odot}$ within a radius of $\sim$0.75\,kpc. This is in good agreement with the total mass of
4.3$\times10^8$~\Msol\ estimated by TIL96 from \hi\ emission features detected
by \citet{Thuan:1995}. It should be noted that all previous kinematical studies of
BCDs have been based on \hi\ 21\,cm observations
\citep[][]{VanZee:2001,Thuan:2004}. 

The broad line component C2 also shows a velocity gradient roughly
perpendicular to the rotation axis, (Fig\,~\ref{fig:velcuts}b), aligned more
closely with the major axis of Mrk~996, with radial velocities that merge with
those of C1 2$''$ from the centre of the galaxy, but extending to larger
negative values. No definite velocity structure can be seen along this axis for
the C3 and C4 components; however, these show an opposing velocity gradient
along the galaxy's rotation axis (Fig\,~\ref{fig:velcuts}a). Their radial
velocities lie much higher at $-$60 to $-$160 km s$^{-1}$ for C3 and $+$110 to
$+$210 km s$^{-1}$ for C4, with the gradients in both showing a similar `S'
shape offset by 1$''$ from the dynamical centre of Mrk~996. It is further noted
that the velocity centre of symmetry of the `S' feature is offset by $+$50\,km
s$^{-1}$ from the heliocentric systemic velocity of Mrk~996 measured above. The
`S' kinematic feature can be attributed to a two-arm spiral structure located
at the nucleus of Mrk~996, whose approaching and receding arms are respectively
traced by the velocity distribution of the C3 and C4 components of H$\alpha$.
Furthermore, our analysis reveals that the projected angular velocity vectors
on the plane of the sky of the spiral and of the gas traced by the C1 narrow
line component are at an angle of $<$90$^\circ$. In their WFPC2 images of
Mrk~996, TIL96 identified a small spiral structure along the east-west
direction (see their fig 4b) whose pivot is the nuclear star-forming region
which is slightly offset from the centre of the outermost F791W image
isophotes. The pivot of the spiral and the peak flux position on the VIMOS
H$\alpha$ C1 map coincide spatially. The 1$''$ offset mentioned above for the
spatial centre of symmetry of the `S' feature (Fig\,~\ref{fig:velcuts}a) is
consistent with the asymmetry noted by TIL96. This fact, along with the
$\sim$$+$50 \,km s$^{-1}$ offset of the spiral's pivot with respect to the
systemic velocity of the galaxy, indicates that the spiral-like nucleus of
Mrk~996 is kinematically decoupled to some degree from the main ionised gas
component. This could indeed be the fossil kinematic signature of a past merger
event.

The FWHMs of C1--4 in H$\alpha$ are shown in Fig\,~\ref{fig:Hamaps}c, corrected
for the instrumental PSF (see Section \ref{sec:reduction}).  A peak in the FWHM
of the narrow C1 component, $\sim$110 km s$^{-1}$, is seen at the nucleus of
Mrk~996.  This gradually, but not uniformly, decreases to $\sim$65 km s$^{-1}$
towards the outer regions. The gradient in the FWHM of C2 is far stronger,
peaking at $\sim$450 km s$^{-1}$ at two positions northwest and southest of the
nucleus and decreasing to $\sim$100 km s$^{-1}$ elsewhere: the broader line
widths along the former direction could be the signature of an expanding shell
surrounding the inner star forming region, blown outwards by the stellar winds
of the core cluster. Higher resolving power IFU spectroscopy is needed to
disentangle the kinematics of this region. No significant gradient is seen in
the FWHM of components C3 and C4, both remaining between 200--300 km s$^{-1}$.

\begin{figure}
\begin{center}
\includegraphics[height=6.5cm,width=8cm]{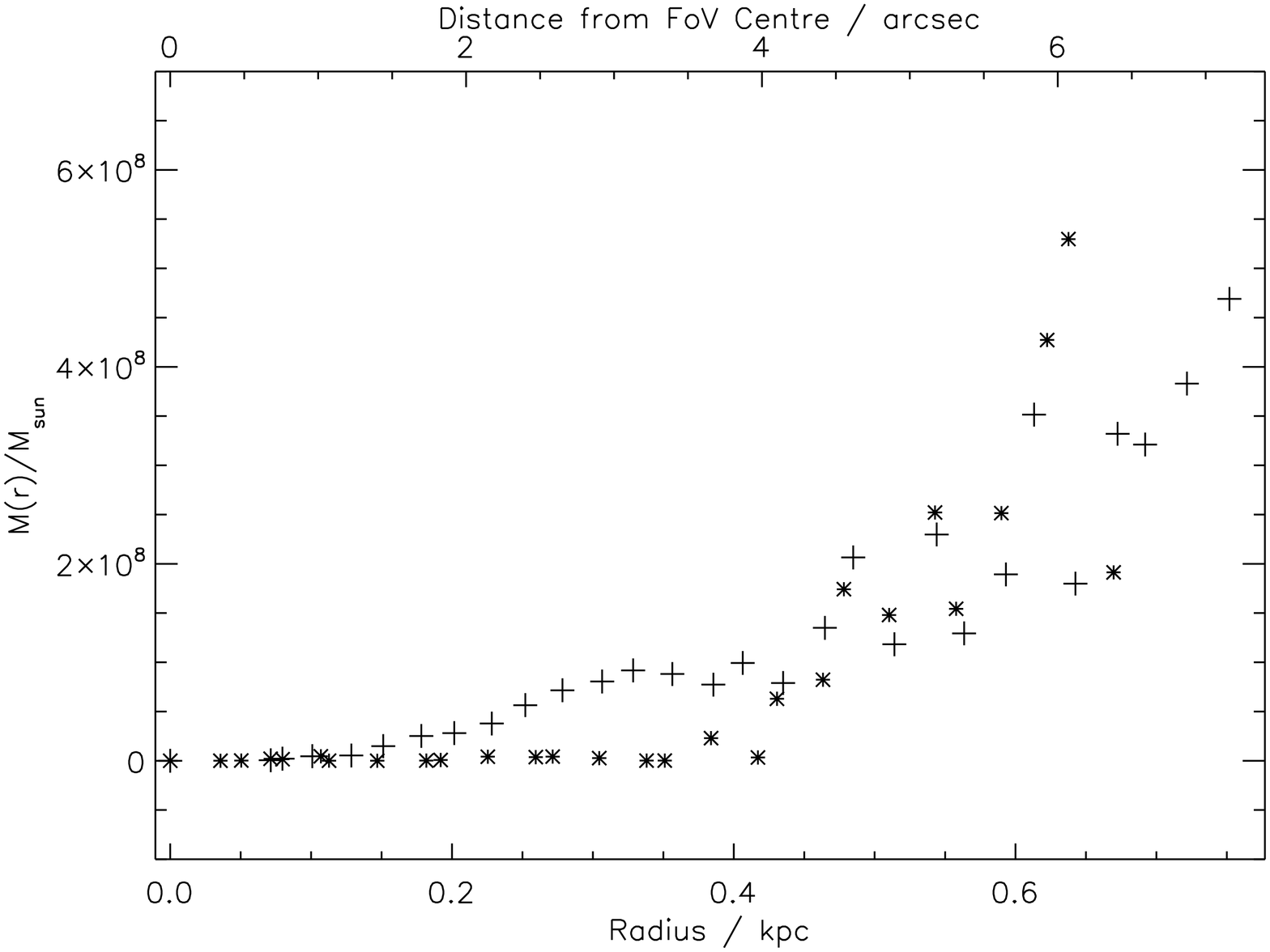}

\caption{Dynamical mass as a function of radius, $r$, derived from the
distribution of C1 radial velocities along the direction normal to Mrk~996's
rotation axis: distance east from the centre represented by crosses and distance west from the centre represented by stars.} \label{fig:dynMass}
\end{center}
\end{figure}

\subsection{The [O~{\sc iii}] nebular line properties}
\label{sec:o3neb}

As mentioned previously, the structure of the \hi\ Balmer lines is not mirrored
in the forbidden nebular lines.  All the strong forbidden lines (with the
exception of [O\three] $\lambda$4363 and [N\two] $\lambda$5755 discussed below)
are comprised of a narrow, central Gaussian (C1) with an underlying {\it
single} broad Gaussian component (C2).  An optimal fit, after applying the
rigorous statistical F-test to a spaxel from the core of Mrk~996, can be seen
in Fig\,~\ref{fig:fit_comps}f. For the spaxels in which the Balmer components
C3 and C4 are detected, the FWHM of the [O~{\sc iii}] $\lambda$5007 C2
component is $\sim$390--415 km s$^{-1}$ and that of each of H$\alpha$ C3 and C4
is $\sim$180--240 km s$^{-1}$.  The occurrence of additional C3-4 components in
the Balmer lines is in contrast to the non-occurrence of C2 splitting in the
[O~{\sc iii}] nebular lines and can perhaps be ascribed to a lower S/N ratio
for the broad component of [O~{\sc iii}] $\lambda\lambda$4959, 5007 compared to
the C2 of \hi\ lines: the observed C2/C1 flux ratios pertaining to the [O~{\sc
iii}] nebular lines in the inner core of Mrk~996 (as defined in the following
subsection) are half of those observed in the \hi\ lines (Table~2). However,
from our analysis of the broad C2 component, we find log (\eld/\cmt) $=$ 7.25
$\pm$ 0.25 from the \foiii\ lines (Section~\ref{sec:broad}). This is high
enough to collisionally suppress $\lambda\lambda$5007+4959 whose critical
density for de-excitation is 6.4$\times$10$^5$ cm$^{-3}$. The distribution of
flux in the C1 and C2 components of the \foiii\ $\lambda$5007 line is shown in
Fig\,~\ref{fig:O3aFluxMap}. The shape and extent of the emitting regions are
similar to those of the corresponding H$\alpha$ C1 and C2 components.

\subsection{Broad emission lines: [O\three] $\lambda$4363 and [N\two] $\lambda$5755}
\label{sec:broads}

\begin{figure*}
\begin{center}
\includegraphics[scale=0.55]{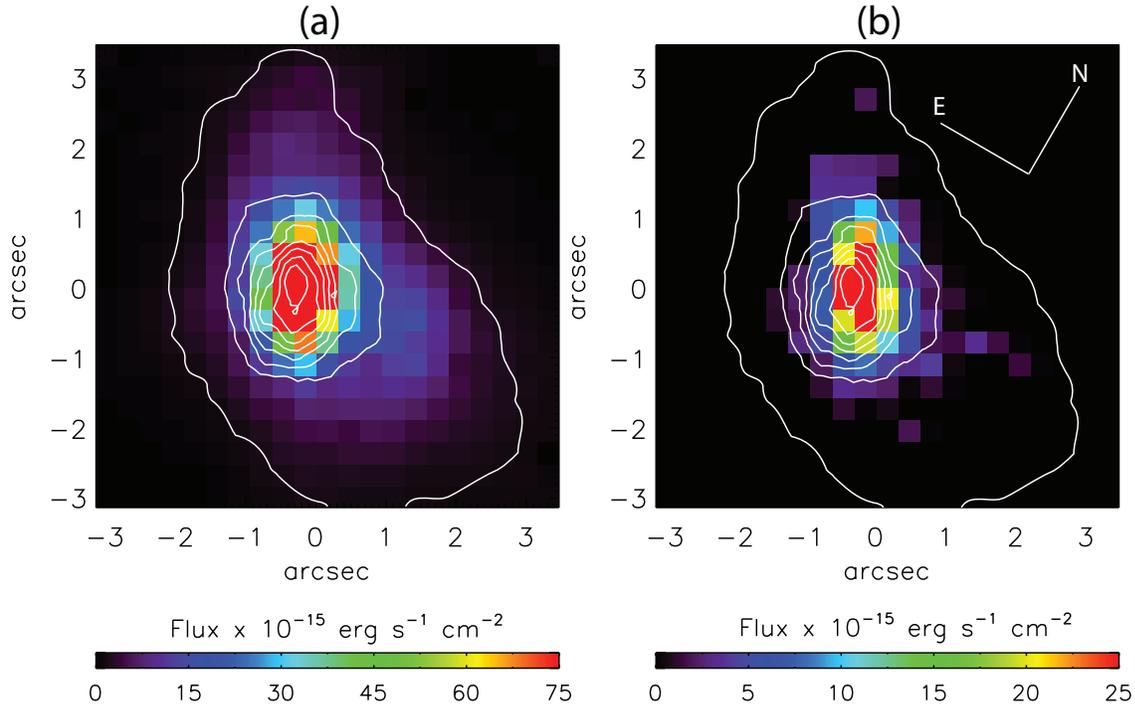}

\caption{[O\three] $\lambda$5007 emission line maps of Mrk~996 showing the flux
distribution per 0.33$\times$0.33 arcsec$^2$ spaxel: (a) in the narrow C1 component;
(b) in the broad C2 component. Overlaid are the H$\alpha$ narrow component
contours (white), as shown in Fig\,~\ref{fig:HSToverlay}.}
\label{fig:O3aFluxMap}
\end{center}
\end{figure*}
\begin{figure}

\begin{center}
\includegraphics[height=8cm,width=8cm]{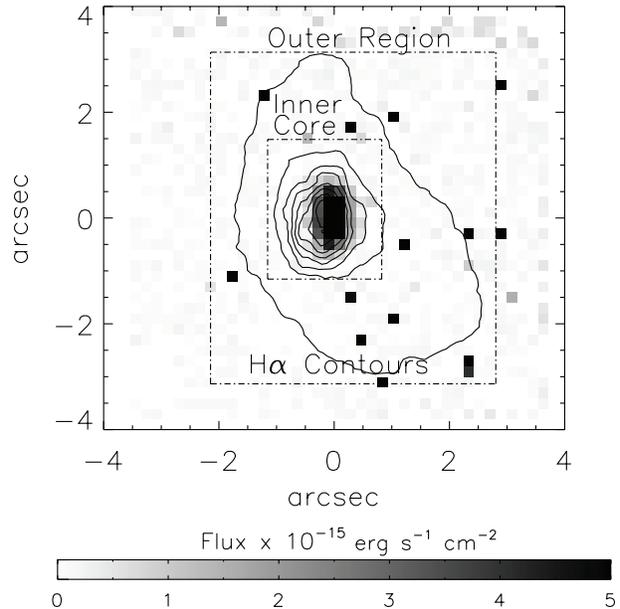}
\caption{Flux distribution of [O\three] $\lambda$4363 per 0.33$\times$0.33
arcsec$^2$ spaxel with H$\alpha$ contours overlaid. The area over which $\lambda$4363
is detected is used to define an `inner core' region, with a surrounding `outer
region' covering the full extent of the H$\alpha$ emission.}
\label{fig:O3cregions}
\end{center}
\end{figure}

Our multi-component line analysis shows that only the broad (C2) components of
the auroral [O\three] $\lambda$4363 and [N\two] $\lambda$5755 transitions are
present, and only in the inner galaxy. The implications of this for the use of
\foiii\ $\lambda$4363 as a temperature diagnostic are discussed in Section
\ref{sec:broad}. For the purpose of this study the $\lambda$4363 spectral map
(Fig\,~\ref{fig:O3cregions}) is used to define a two-region model of Mrk~996:
an `inner core' region, defined by $\lambda$4363 emission extending over an
area of 1.7$\times$2.3 arcsec$^{2}$ (185$\times$250 pc$^2$), and an `outer
region' large enough to incorporate almost the full extent of H$\alpha$
emission over an area of 5.3$\times$6.3 arcsec$^2$ (575$\times$685 pc$^2$). The
lack of any detectable \foiii\ $\lambda$4363 emission in the outer region is
illustrated in Fig\,~\ref{fig:residuals4363}. Here we compare the summed
spectrum of the outer region plus core with the summed spectrum of the core
region, and also show the residual outer region spectrum. An emission line map
of the broad \fnii\ $\lambda$5755 line is shown in Fig\,~\ref{fig:N2bc}a; its
flux distribution correlates strongly with that of the similarly broad
$\lambda$4363, suggesting that the excitation of these lines is affected by
processes that are particular to the inner core region of Mrk~996. In
Fig\,~\ref{fig:allbroad} the velocity profiles of the broad C2 components of
H$\alpha$, H$\beta$ and [O\three] $\lambda$5007 are shown, overlaid in velocity
space with the profiles of the [O\three] $\lambda$4363 and [N\two]
$\lambda$5755. Each spectral line has been summed over the inner core region
and all show a FWHM of $\sim$300 kms$^{-1}$, suggesting that a common
excitation mechanism is responsible for the emission of the broad component.
The larger widths of [O\three] $\lambda$4363 and [N\two] $\lambda$5755 indicate
that the physical conditions within the inner core are different from those in
the outer region.

\begin{figure}
\begin{center}
\includegraphics[height=6cm,width=8cm]{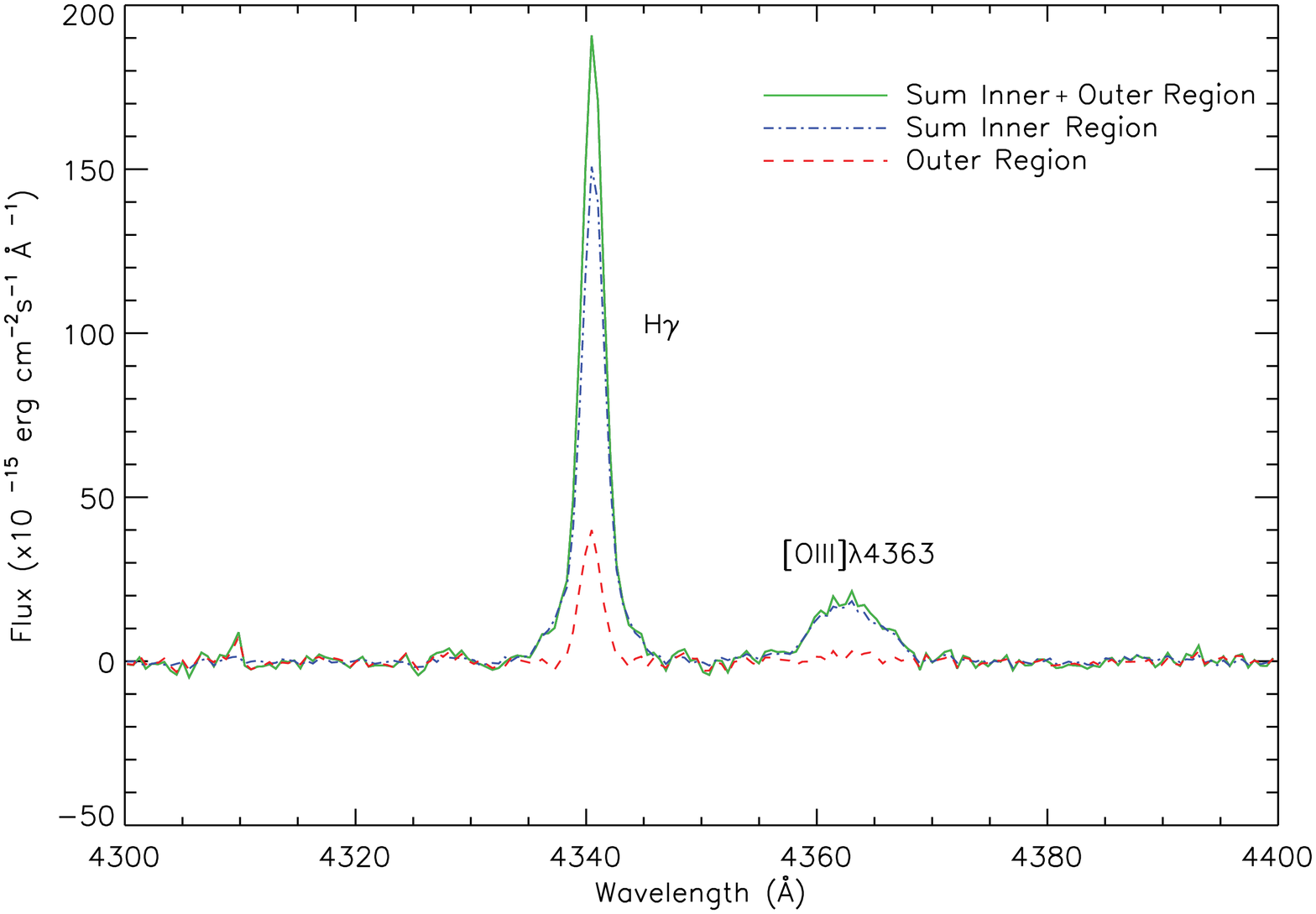}

\caption{Summed spectra from the inner core and the outer region illustrating
the lack of detectable \foiii\ $\lambda$4363 in the outer region of Mrk~996.}
\label{fig:residuals4363}
\end{center}
\end{figure}

\begin{figure*}
\begin{center}
\includegraphics[scale=0.5]{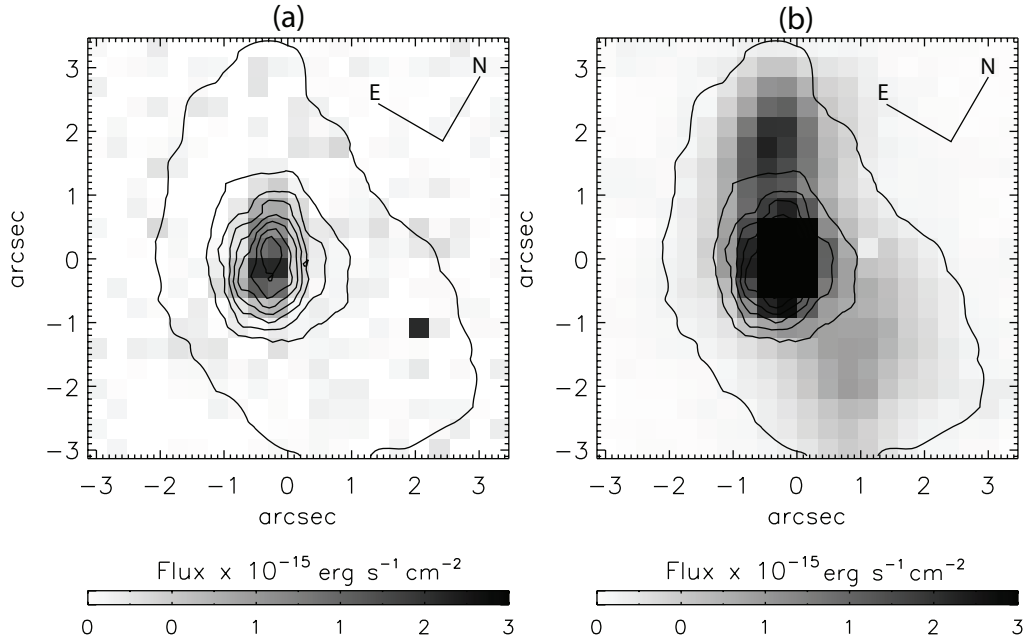}

\caption{Flux distribution per 0.33$\times$0.33 arcsec$^2$ spaxel of (a) [N\two]
$\lambda$5755 for which only a broad component is detected and which is seen
only within the core region; (b) [N\two] $\lambda$6584, whose flux distribution
peaks in the core and towards the north-east; a fainter peak is seen west of
the core. Overlaid are the H$\alpha$ narrow component contours as shown in
Fig\,~\ref{fig:HSToverlay}.} \label{fig:N2bc}
\end{center}
\end{figure*}

\begin{figure}
\begin{center}
\includegraphics[height=6.5cm,width=8cm]{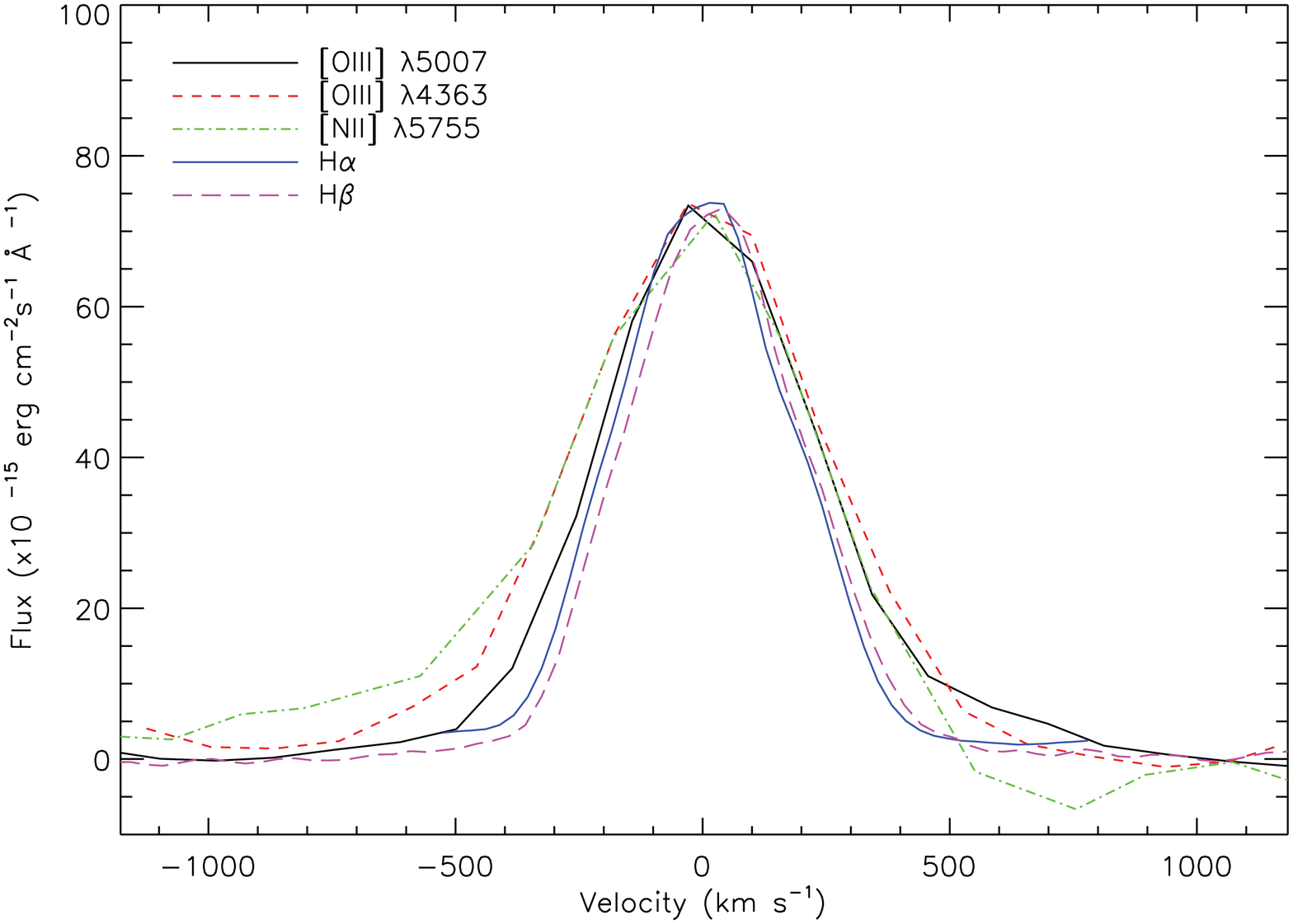}

\caption{Velocity profiles of the broad C2 components of H$\alpha$, H$\beta$,
\foiii\ $\lambda$5007, and of \foiii\ $\lambda$4363, \fnii\ $\lambda$5755
emission lines (relative to the heliocentric systemic velocity of $+$1642 km
s$^{-1}$). Spectra have been summed over the inner core 1.7$\times$2.3
arcsec$^2$ region of Mrk~996.} \label{fig:allbroad}
\end{center}
\end{figure}

\section{Multi-Component diagnoses: electron temperature and density}
\label{sec:broad}

The multi-component nature of the emission lines from Mrk~996 must be taken
into account when applying plasma diagnostic methods. For example, since
[O\three] $\lambda$4363 is only detected from the central region of Mrk~996 and
completely lacks the narrow line component that [O\three] $\lambda\lambda$4959,
5007 exhibit (see Fig\,~\ref{fig:residuals4363}), an integrated [O\three]
($\lambda$5007+$\lambda$4959)/$\lambda$4363 ratio can not provide a useful
temperature diagnostic; this would lead to incorrect results. Previous analyses
lacked the spectral and spatial resolution needed to decompose the line
profiles and were forced to assume a single density of \eld\ $\sim$10$^6$
cm$^{-3}$ in order to match the anomalously low integrated [O\three]
($\lambda$5007+$\lambda$4959)/$\lambda$4363 flux ratio (TIL96).

The detection of separate narrow and broad line components opens up the
possibility of determining separate electron temperatures and densities and
conducting a separate abundance analysis for each.

\subsection{The Narrow Component Gas}
\label{sec:narrowTeNe}

In order to estimate the electron temperature (\elt) from which the narrow line
component C1 emission arises the following method was used. An upper limit to
the \elt\ applicable to the \hi\ and [O\three] nebular C1 components can be
obtained via the simulation of a narrow \foiii\ $\lambda$4363 component using
the observed narrow $\lambda$5007 component. Predicted $\lambda$4363 C1 line
intensities were obtained by multiplying the observed $\lambda$5007 C1 profile
by the theoretical $\lambda$4363/$\lambda$5007 intensity ratio for a range of
\elt's at \eld\ $=$ 170 cm$^{-3}$; the latter was measured from the [S\two]
doublet ratio as described below. It should be noted that the \fsii\ lines only
show narrow C1 components (Table~\ref{tab:fluxvals}). The scaling factors were
applied to both the inner core and outer region C1 component $\lambda$5007
absolute fluxes. Simulated narrow $\lambda$4363 profiles of decreasing
intensity were then subtracted from the observed $\lambda$4363 line until the
residual was no longer detectable, as shown for the core region $\lambda$4363
profile in Fig\,~\ref{fig:upperTe}. It is found that a measurable narrow
$\lambda$4363 C1 component would be detectable at an electron temperature of
$>$10,000~K for both the inner core and the outer region. Thus we adopt an
upper limit of 10,000~K to the electron temperature of the gas emitting the C1
component throughout Mrk~996.  In Section~\ref{sec:broad_abunds} we argue that
this limit is closer to 9200~K, based on the inference that these O/H abundance
ratio across the narrow and broad line emitting regions is invariant.

\begin{figure}
\begin{center}
\includegraphics[height=5.5cm,width=7cm]{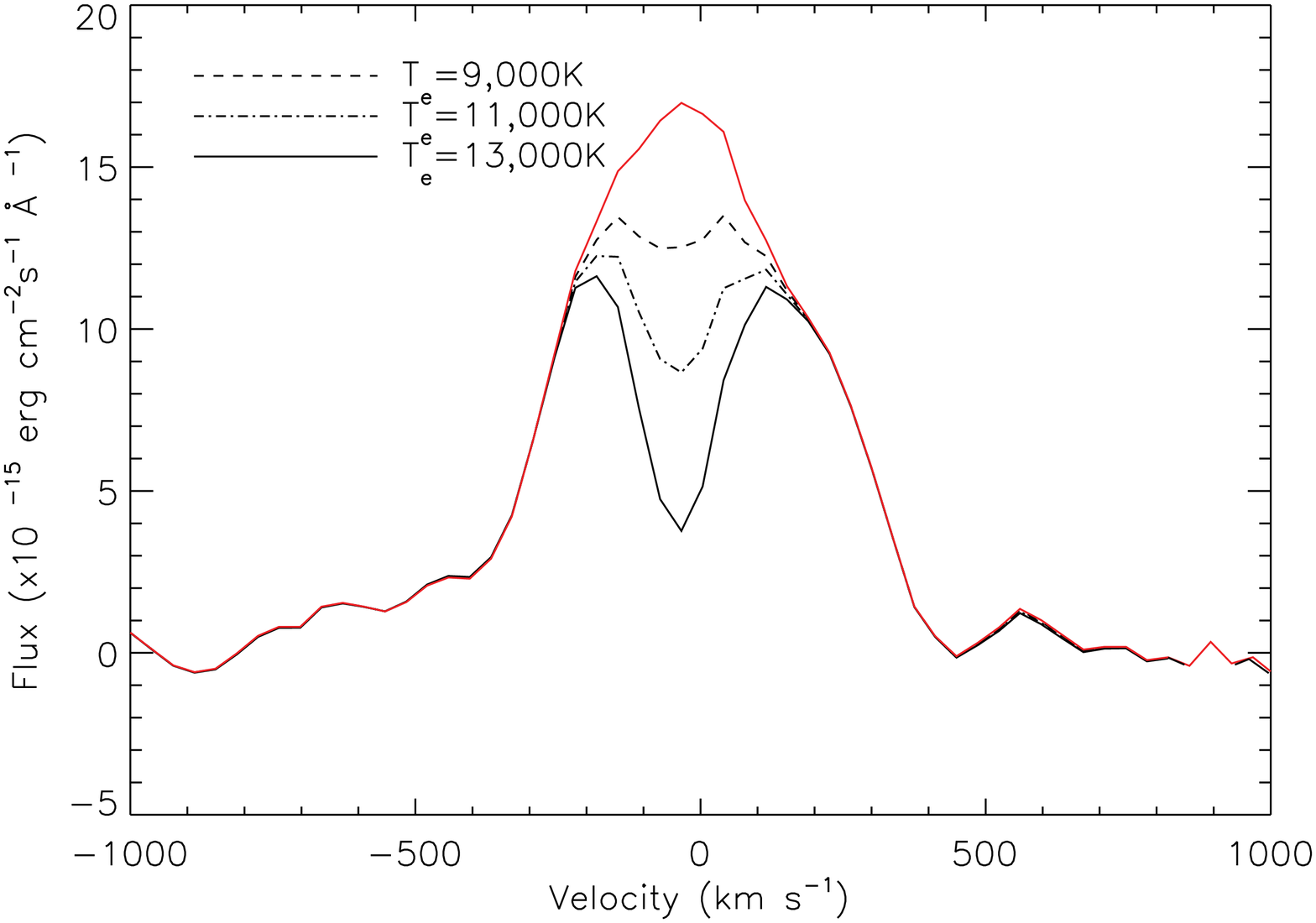}

\caption{Inner core region dereddened [O\three] $\lambda$4363 profile (solid red curve) with
simulated narrow components subtracted from it, corresponding to the dereddened [O\three]~$\lambda$5007 narrow component profile, normalised to the different $\lambda$4363 flux levels that correspond to different adopted T$_e$ values. The subtracted narrow component
profiles correspond to electron temperatures of 9000~K, 11,000~K and 13,000~K for an
electron density of 170 cm$^{-3}$. An upper limit of 10,000~K is estimated for the \elt\ of the gas emitting the narrow C1 component in the inner core region. See text for details.}
\label{fig:upperTe}
\end{center}
\end{figure}

\begin{table}
\centering
\caption{Electron densities derived from the narrow line [S\two]
$\lambda$6716/$\lambda$6731 intensity ratios (1.26$\pm$0.03 and 1.41$\pm$0.03,
respectively for the inner core and outer region of Mrk~996).}

\begin{tabular}{ccc}
\hline
Adopted \elt\ (K) & \eld\ (cm$^{-3}$) & \eld\ (cm$^{-3}$)  \\
&Core Region & Outer Region\\

\hline
9,000 & 170$\pm$40 & 10$\pm^{30}_{9}$  \\
11,000 & 170$\pm$40 & 10$\pm^{30}_{9}$  \\
13,000&  170$\pm$40&  10$\pm^{30}_{9}$ \\
\hline
\end{tabular}
\label{tab:SIINe}
\end{table}

The [S\two] $\lambda\lambda$6717, 6731 lines, whose intensity ratio is a common
electron density diagnostic for \hii\ regions, show only the narrow emission
component and were used to compute density values for a range of electron
temperatures for the inner and outer regions of the galaxy
(Table~\ref{tab:SIINe}, using the {\sc iraf}'s\footnote{{\sc iraf} is
distributed by the National Optical Astronomy Observatory, which is operated by
the Association of Universities for Research in Astronomy} {\sc temden} task of
the {\sc nebula} package). These densities are representative of conditions in
the gas from which the narrow component emission arises. This propels us to use
alternative methods to investigate the density of the gas emitting the broad
line components in the inner core region of Mrk~996. With the electron
temperature of \elt\ $\le$ 10,000~K obtained above, we find that electron
densities of 170 cm$^{-3}$ and 10 cm$^{-3}$ are representative of the narrow C1 component
gas in the inner and outer regions of Mrk~996, respectively.

\subsection{[Fe\three] line diagnostics}
\label{sec:FeIII}

\ffeiii\ emission line ratios provide useful density diagnostics that can trace
both high and low electron densities; see \citet{Keenan:2001}. The summed inner
region spectrum of Mrk~996 (Fig\,~\ref{fig:fullSpec}) shows a few of the brightest transitions among the
3d$^6$ levels of Fe~{\sc iii}; [Fe\three] $\lambda\lambda$4658, 4702, 4881,
4986, and 5270. The flux in these lines above the continuum level was measured
by integrating over the line profile. To maximise accuracy, the fluxes for
$\lambda$4658 and $\lambda$4702 were measured after removing any contamination
from Wolf-Rayet stellar features by subtracting the WR template fit derived in
Section~\ref{sec:WR}. Fig\,~\ref{fig:FeIIIdens} shows the extinction corrected
fluxes of $\lambda$4881 and $\lambda$5270 relative to $\lambda$4658 along with
theoretical ratios as a function of \eld, for \elt's between 7,000--20,000~K,
from \citet{Keenan:2001}. Interpolating between the theoretical data for an
adopted \elt\ of 10,000~K, the $\lambda$4881/$\lambda$4658 ratio indicates log
(\eld/\cmt) $=$ 5.7 $\pm$ 0.3, which falls within the somewhat larger range of
log (\eld/\cmt) $=$ 6.5$^{+0.5}_{-1.5}$ implied by the observed
$\lambda$5270/$\lambda$4658 ratio. These results indicate the presence of a
very dense ionised medium throughout the inner core of Mrk~996.

\begin{figure}
\begin{center}
\includegraphics[scale=0.32]{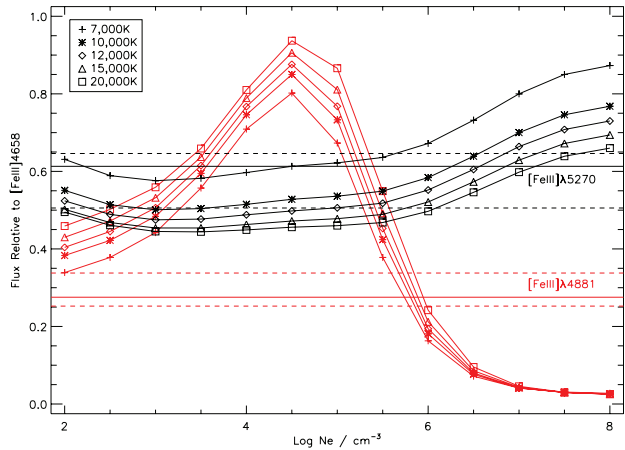}

\caption{Theoretical [Fe\three] line flux ratios, for $\lambda$4881 (red) and
$\lambda$5270 (black), relative to $\lambda$4658 as a function of electron
density for temperatures between 7,000 -- 20,000~K. The horizontal lines denote
the observed flux ratios (solid) and their uncertainties (dashed) from the
summed spectra of the inner core region of Mrk~996.} \label{fig:FeIIIdens}
\end{center}
\end{figure}

\begin{figure}
\begin{center}
\includegraphics[scale=0.32]{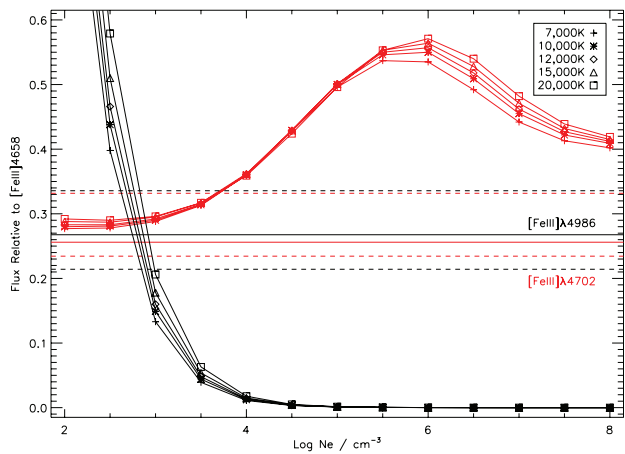}

\caption{Theoretical [Fe\three] line flux ratios, for $\lambda$4986 (black) and
$\lambda$4207 (red), relative to $\lambda$4658 (in the region in which $\lambda$4986 is
seen), relative to $\lambda$4658 as a function of electron density for
temperatures between 7,000--20,000~K.   The horizontal lines denote the
observed flux ratios (solid) and their uncertainties (dashed) from the summed
spectra of the region where $\lambda$4986 is detected.}
\label{fig:FeIIIdens4986}
\end{center}
\end{figure}

In contrast to the above however, the [Fe\three] $\lambda$4986/$\lambda$4658 ratio is a
more sensitive tracer of low electron densities. Fig\,~\ref{fig:FeIIIdens4986}
shows that relative to $\lambda$4658 the $\lambda$4986 line should not be
detectable at densities higher than $\sim$1,000 cm$^{-3}$; the fact that the
line is detected from Mrk~996 indicates the presence of a low density gas
component as well. This is not inconsistent with the high densities derived
above for the inner core as the spatial distribution of $\lambda$4986 emission
(not shown) contrasts strongly with that of the other [Fe\three] lines --
rather than peaking in the core of the galaxy, $\lambda$4986 is distributed
throughout Mrk~996 and is not confined to the inner region defined by the broad
[O\three] $\lambda$4363 emission. There is a clear detection of $\lambda$4986
emission over a group of 8 spaxels extending $\sim$1$''$.6 east from the core.
The $\lambda$4986/$\lambda$4658 and $\lambda$4702/$\lambda$4658 intensity
ratios integrated over these 8 spaxels (Fig\,~\ref{fig:FeIIIdens4986}) confirm
that this is a low density area (\eld$\sim$100--1000 cm$^{-3}$), typical of
normal \hii\ regions. This emission could be tracing lower density ionised gas
throughout Mrk~996. In a study of
high excitation nebulae in the Magellanic Clouds by \citet{Naze:2003}, \ffeiii\
$\lambda$4986 was only detected in the shell of the super-bubble surrounding
the bright H$\alpha$ nebula N44C. Due to the large distance of Mrk~996, we are
unable to resolve any bubbles surrounding the central super-star cluster even
though their presence in the galaxy is likely.

\subsection{The Broad Component Gas}
\label{sec:broadTeNe}

The [O\three] ($\lambda$5007+$\lambda$4959)/$\lambda$4363 intensity ratio can
be used to determine the electron temperature of the gas from which the broad
component emission arises. Since, however, \foiii\ $\lambda$4363 is only
detected as a broad line, this dictates that \elt\ estimates using [O\three] can
only involve the broad C2 components; moreover these will only be
applicable to the inner core region of the galaxy where $\lambda$4363 is
detected. The FWHMs of components C2 in $\lambda\lambda$4959, 5007 are
very similar to that of $\lambda$4363, all being in the range of 320--515
kms$^{-1}$ across the inner core region; their surface brightness distributions
are also similar. Also in support of the suitability of the [O\three] ratio
involving the C2 components is that we do not find any indication of C3 or C4
components in the highest S/N emission line profiles of $\lambda\lambda$4959,
5007 (see Fig\,~\ref{fig:fit_comps}(f) for an optimised two-component fit to
the $\lambda$5007 emission line from the inner core region).  Even so, electron
temperatures derived using only the broad line components are un-physically
high ($>$30,000~K) until densities higher than \eld\ $=$ 10$^6$ cm$^{-3}$ are adopted.

We can attempt to constrain the electron temperature and density of the broad
line component by using the $\lambda$1663/$\lambda$4363 and
$\lambda$5007/$\lambda$4363 intensity ratios, which due to their considerably
different O$^{2+}$ upper level excitation energies can act as sensitive
temperature and density diagnostics between the critical densities of each of
these lines.  The FOS UV spectra presented by TIL96 revealed a weak O\three]
$\lambda$1663 line detection which, due to its high excitation energy ought to
originate from the same gas emitting the broad [O\three] $\lambda$4363 that we
detect. Using a simulated FOS aperture on our broad component flux maps for
$\lambda$5007 and $\lambda$4363, fluxes of 39.6$\pm$6.3$\times10^{-15}$ ergs
cm$^{-2}$ s$^{-1}$ for $\lambda$4363 and 214.1$\pm$14.2$\times10^{-15}$ ergs
cm$^{-2}$ s$^{-1}$ for $\lambda$5007 are predicted to fall within the small FOS
aperture centred on the nucleus of Mrk~996. Together with the $\lambda$1663
flux of 1.9$^{+0.4}_{-0.2}$$\times10^{-15}$ ergs cm$^{-2}$ s$^{-1}$ measured
from the FOS spectrum, these fluxes were then dereddened using an average
$c$(H$\beta$) of 0.65 for a simulated FOS aperture on the broad component VIMOS
$c$(H$\beta$) map, to yield $\lambda$4363/$\lambda$1663 and
$\lambda$5007/$\lambda$4363 intensity ratios of 2.4 and 4.3, respectively.  The
theoretical $\lambda$4363/$\lambda$1663 and $\lambda$5007/$\lambda$4363 ratios
for log(\eld) $=$ 5.0--10.0 and \elt\ $=$ 5,000--30,000~K are presented in
Fig\,~\ref{fig:1663_plots}a, b and the resultant set of solutions are shown in
Fig\,~\ref{fig:1663_plots}c.  The two curves in Fig\,~\ref{fig:1663_plots}c
overlap between log(\eld) $=$ 6.5--8.5 and \elt\ $=$ 7,000--14,000~K.  We thus
adopt \elt\ $=$ 10,500 $\pm$ 3,500~K and log(\eld/cm$^{-3}$) $=$ 7.25$^{+1.25}_{-0.75}$
for the broad component emitting gas. The inferred density range is somewhat
higher than the value derived from the [Fe\three] $\lambda$4881/$\lambda$4658
ratio (Section~\ref{sec:FeIII}) and supports the conclusion that the central
regions of Mrk~996 contain very dense zones of ionised gas.

\begin{figure*}
\begin{center}
\includegraphics[scale=0.7]{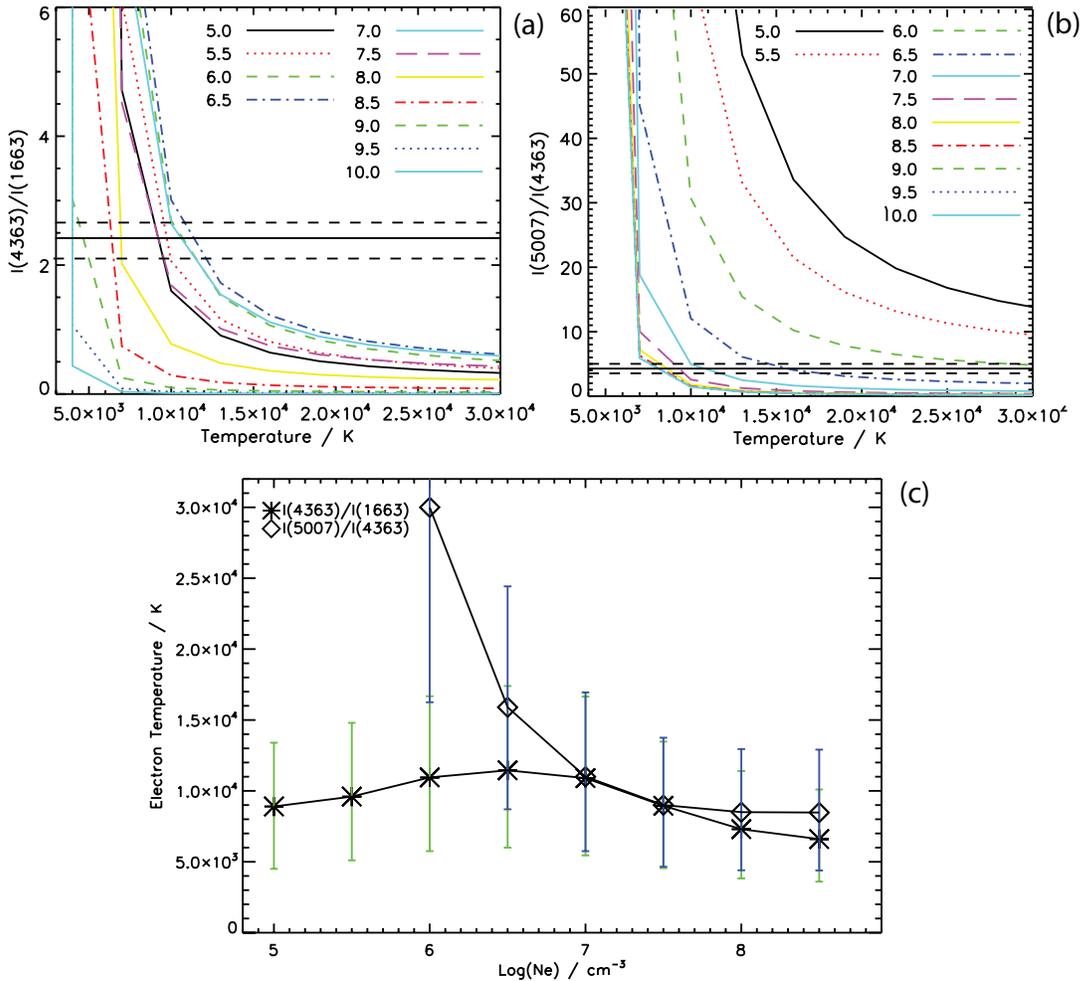}

\caption{Broad component \elt\ and \eld\ diagnostic diagrams for (a)
$I$($\lambda$4363)/$I$($\lambda$1663), and (b)
$I$($\lambda$5007)/$I$($\lambda$4363), showing theoretical ratios for \elt\ $=$
5,000--30,000~K and log(\eld/\cmt) $=$ 5.0--10.0. The horizontal solid lines in
(a) and (b) denote the observed dereddened ratios and their uncertainties (dashed); (c) shows the (\elt, \eld)
solutions applicable to Mrk~996 obtained from (a) and (b) and their uncertainties.}
\label{fig:1663_plots}
\end{center}
\end{figure*}

\section{Chemical Abundances}
\subsection{Narrow Component Abundances}
\label{sec:abund_nar}

\begin{figure*}
\begin{center}
\includegraphics[scale=0.65]{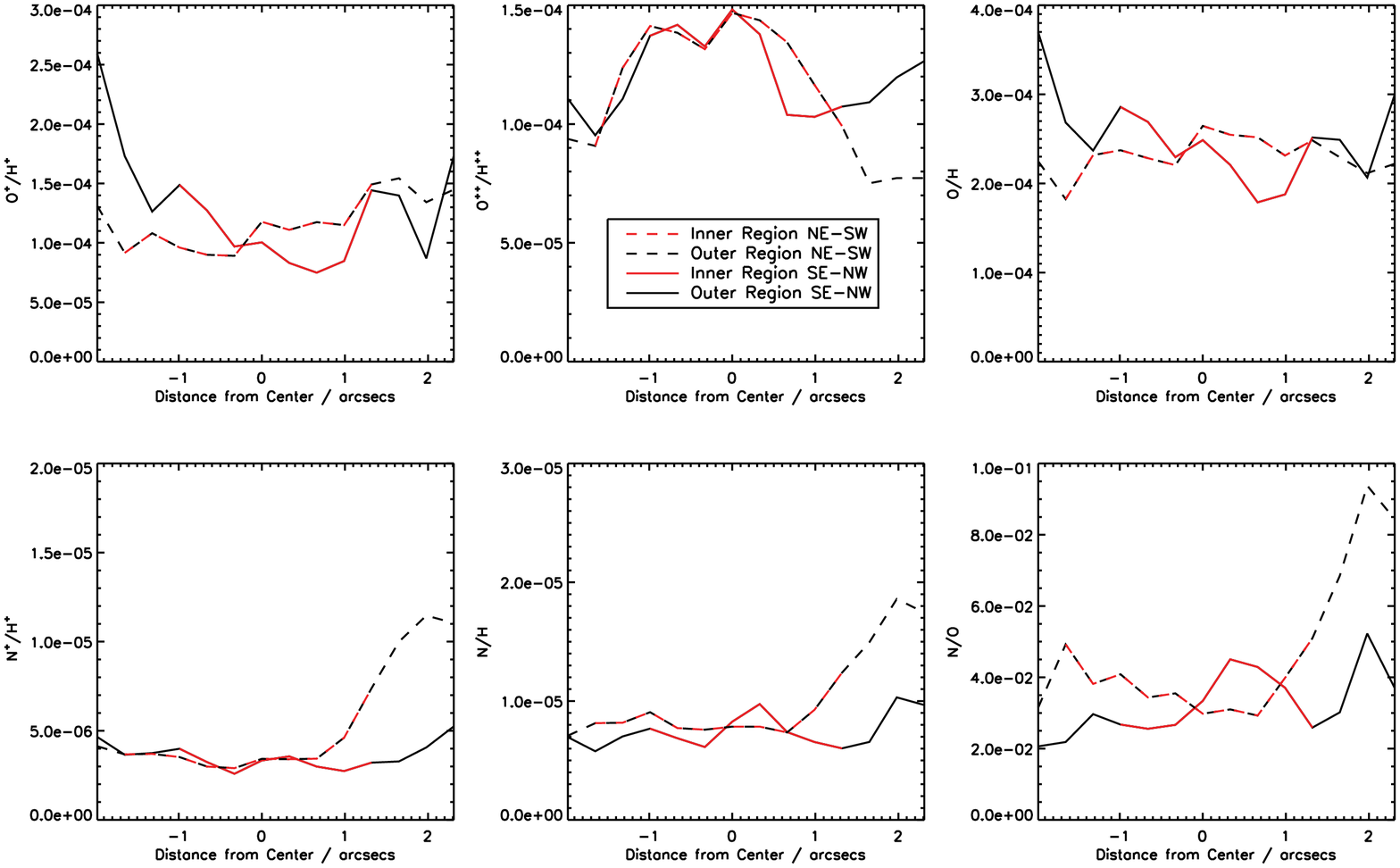}

\caption{Spatial variations in narrow component ionic and elemental abundances
across Mrk~996 obtained for \elt\ $=$ 10,000~K and \eld\ $=$ 170 cm$^{-3}$. The
available \elt\ is an upper limit so the abundances shown are lower limits. The
patterns should persist if \elt\ is constant across the emitting region. Three
spaxel-wide averages were made across the IFU aperture in its $X$ and $Y$
directions, corresponding to south-east to north-west and north-east to
south-west, respectively.} \label{fig:abundcuts}
\end{center}
\end{figure*}

\begin{figure}
\begin{center}
\includegraphics[scale=0.35]{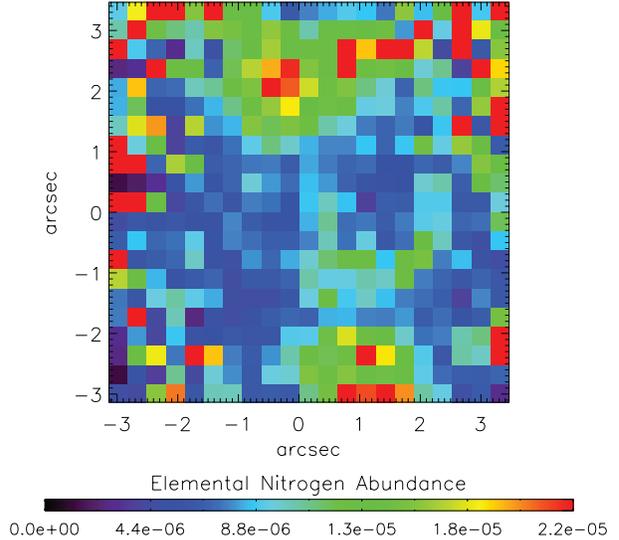}
\caption{N/H abundance ratio map based on the narrow component emission across
the region where \fnii\ $\lambda$6584 is seen.  It should be noted that these
abundances are lower limits since an upper limit to \elt\ was used for the
calculation.} \label{fig:N_H_abund}
\end{center}
\end{figure}

Abundance maps relative to H$^+$ were created for the N$^{+}$, O$^{+}$,
O$^{2+}$, S$^{+}$, S$^{2+}$, Ar$^{2+}$ ions, using the narrow
$\lambda\lambda$6584, 7320+7330, 5007, 6717+6730, 6312, and 7136-\AA\ lines
respectively (via the {\sc abund} task in {\sc iraf}). Each VIMOS spaxel was
treated as a distinct `nebular zone' with its own set of physical conditions.
Ionic argon and sulphur abundance ratios were converted to total Ar/H and S/H
using an ICF(Ar) $=$ 1.87 and ICF(S) $=$ 1.03 respectively \citep{Barlow:1994}.
Very small changes were found by adopting the ICF prescriptions of Izotov et
al. (2006). Since \elt\ maps are unattainable for the narrow component gas, we
have adopted the upper limit of \elt\ $=$ 10,000~K applicable throughout
Mrk~996 (Section~\ref{sec:narrowTeNe}), together with the mean electron
densities derived from the [S\two] doublet ratios, of 170 cm$^{-3}$ and 10
cm$^{-3}$ for the inner core and outer region, respectively
(Table~\ref{tab:SIINe}). It is emphasised that since the adopted \elt\ for the
narrow component gas is an upper limit, the abundances derived here are
necessarily lower limits.  

 In Fig\,~\ref{fig:abundcuts} abundance distributions
are shown in the directions from north-east to south-west and south-east to
north-west that pass through the centre of Mrk~996 (i.e. in the $X$ and $Y$
directions across the abundance maps, respectively, produced by averaging over
a three spaxel-wide pseudo-slit). The variations in the abundances of O$^{2+}$
and O$^{+}$ in Fig\,~\ref{fig:abundcuts} mirror each other in their
distribution. As expected, the more highly ionised species is dominant in the
core region, where the UV radiation is expected to be harder. It should further
be noted that these abundance trends are only correct if the temperature is
constant throughout the gas emitting the narrow line component.

The O/H elemental ratio was derived by summing the O$^{+}$ and O$^{2+}$ ionic
abundance maps. Adopting a solar oxygen abundance of 4.9$\times$10$^{-4}$
relative to hydrogen \citep{Prieto:2001}, we find that Mrk~996 has an oxygen
abundance of $\geq$0.50 solar  throughout the inner core region and
beyond. This is a significantly higher metallicity than the $\sim$0.2 solar derived previously by TIL96, which 
can be attributed to the high electron temperature of 15,000~K that they 
employed, based on {\sc cloudy} models.

An N/H abundance ratio map (Fig\,~\ref{fig:N_H_abund}) was created using an
ionisation correction factor, ICF(N) $=$ (O$^+$+O$^{2+}$)/O$^{+}$
\citep[][]{Barlow:1994}. Again, very small changes were found by adopting the
ICF prescriptions of \citet{Izotov:2006}.  For a constant electron 
temperature, the N/H ratio, shown in
Fig\,~\ref{fig:N_H_abund}, remains constant throughout the core at 
$\geq0.8\times10^{-5}$ but shows a two-fold increase north-east of the core
region of Mrk~996. A cut across the N/O ratio map of Mrk~996, which is 
much less sensitive to the adopted electron temperature than either O/H or 
N/H, shows a similar trend, with log(N/O) $\simeq$ $-$1.5 in the core, 
rising to $\simeq$ $-$1.2 at a spot north-east of the core 
(Fig\,~\ref{fig:abundcuts}).

The mean abundances are listed in Table~5. For the revised oxygen metallicity
of Mrk~996 the derived N/O, S/O and Ar/O ratios fall within the expected range
for BCDs \citep[e.g.][]{Izotov:2006}.

\begin{table*}

\caption{Intensities of UV emission lines from {\it HST} FOS spectra and their
optical broad component counterparts, measured within a FOS aperture
superimposed on our broad component flux maps.  All intensities are given
relative to the broad component $F$(H$\beta$) $=$ 122.1 $\pm$
11.1$\times$10$^{-15}$ erg s$^{-1}$ cm$^{-2}$ within the simulated aperture.}

\begin{center}
 \begin{tabular}{lc|lccc}
  & & &\multicolumn{3}{c}{Abundance $\times$10$^4$}\\
  \cline{4-6}
  & &T$_e$ / K& 9,000 & 10,000 & 11,000 \\
& & log(N$_e$/cm$^{-3}$) & 7.50 & 7.25 & 7.00 \\
\hline
Line & 100$\times$I($\lambda$)/I(H$\beta$) & Species & & & \\
\hline
 5755 [N\two] & 7.1$\pm$1.9  & N$^{+}$/H$^{+}$& 1.9$\pm$0.5 & 0.9$\pm$0.2 & 0.4$\pm$0.1 \\
 1750 N\three] & 19.2$\pm$6.8& N$^{2+}$/H$^{+}$ & 7.5$\pm$2.6 & 3.0$\pm$1.1 & 1.3$\pm$0.5 \\
 \cline{4-6}
  & & N/H & 9.4$\pm$1.3 & 3.8$\pm$1.7 & 1.7$\pm$0.8 \\
5007 [O\three] &51.3$\pm$6.2 &O$^{2+}$/H$^{+}$& 12.6$\pm$1.5 & 4.8$\pm$0.6 & 2.0$\pm$0.2 \\
\cline{4-6}
& & O/H & 15.7$\pm$7.5 & 5.5$\pm$2.6 & 2.3$\pm$1.1 \\
\hline
\end{tabular}
\label{tab:broadAbund}
\end{center}
\end{table*}

\subsection{Broad component abundances}
\label{sec:broad_abunds}

\begin{figure}
\begin{center}
\includegraphics[scale=0.30]{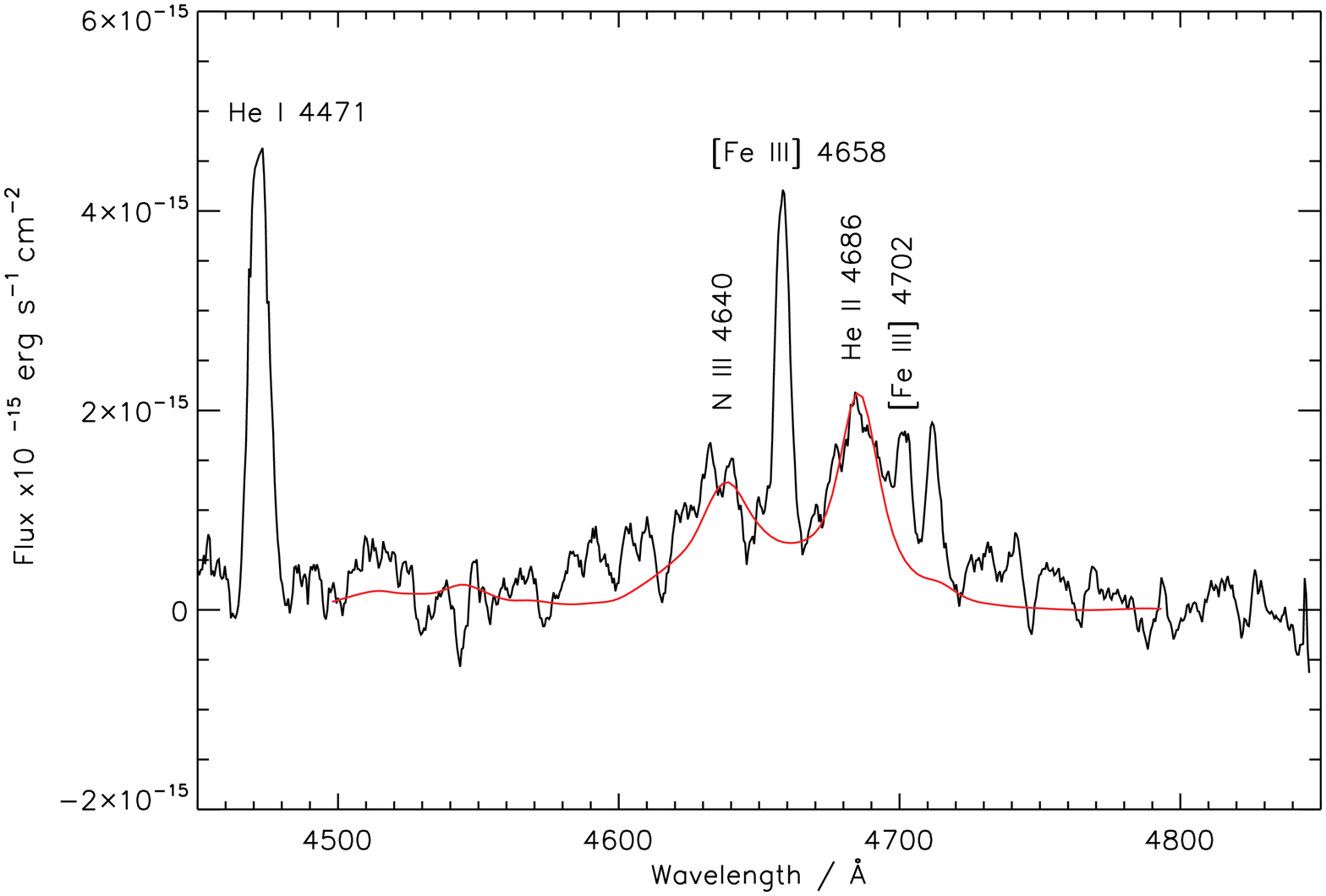}
\caption{WR `blue bump' feature summed over the inner core of Mrk~996.
Overplotted is the combined fit for 2600 WNL stars and 400 WC stars, using the
templates of \citet{crowther:2006}.} \label{fig:WRfit}
\end{center}
\end{figure}

In order to derive nitrogen elemental abundance for the broad component gas, we
correct for the presence of N$^{2+}$ using the N\three] $\lambda$1750 line
detected by {\it HST} FOS. This high excitation energy UV line was not used to
correct the narrow emission component nitrogen abundances because it ought to
originate from the high excitation gas responsible for the \emph{broad} line
emission. Table~\ref{tab:broadAbund} lists the ionic abundances derived for the
broad line emission gas derived from optical and UV lines. Fluxes are quoted
relative to the broad H$\beta$ flux within the simulated FOS aperture and are
corrected for extinction using the average $c$(H$\beta$) value within that
aperture in the reddening map derived from the broad components of
H$\alpha$/\hb, H$\gamma$/\hb\ ratios.  The physical conditions adopted for
these calculations are those derived from the broad component emission, i.e.
\elt\ $=$ 10,000 $\pm$ 3,500~K and log(\eld/cm$^{-3}$) $=$ 7.25 $^{+1.25}_{-0.75}$.  We
consider that the highest allowed \elt\ is 11,000~K with log(\eld/cm$^{-3}$) $=$
7.00 being the lowest allowed density, based on Fig\,~15c.  This latter density
is consistent with the upper density limit inferred from the \ffeiii\
diagnostics.  For \elt\ $=$ 11,000~K and log(\eld/cm$^{-3}$) $=$ 7.00 
 the resulting broad-line O/H abundance ratio of $2.3\times10^{-4}$ is 
consistent with being the same as for  the narrow line emission
regions (cf. Tables~\ref{tab:broadAbund} and \ref{tab:abundSummary}). As
a consistency check, we determined the \elt\ for the narrow component
VIMOS \foiii\ $\lambda$5007 and \hb\ fluxes falling within the FOS 
aperture that would be required to match the above broad-line O/H ratio. This
yielded a temperature of $\sim$9,200~K, reinforcing the upper limit of 10,000~K
derived in Section~\ref{sec:narrowTeNe} for the C1 emission. S/H abundances
were computed using the [S\three]~$\lambda$6312 line and an estimated (minor)
contribution from S$^+$/H$^+$ (in parentheses in Table~5) using equation A38 of
\citet{Barlow:1994}. Ar/H abundance ratios were derived using the
[Ar\three]~$\lambda$7136 line with ICF(Ar)=1.87 as previously. The adopted mean
values are listed in Table~5.

In contrast to the apparent lack of significant variation in the 
abundances of oxygen, sulphur, and argon between the two velocity 
components, it appears that  the N/O ratio in the broad line region is 
enriched by a factor of $\sim$20 relative to the narrow line region, with 
the broad line region having log(N/O) $\sim$ $-$0.13. We 
conclude that the observations are consistent with the narrow and broad 
line regions both having an O/H ratio of $\geq$0.5 solar.

\subsection{Elemental Helium Abundance}

Narrow and broad component He$^{+}$/H$^+$ abundance maps were derived from
\hei\ $\lambda$5876 C1 and C2 maps. Abundances were calculated using the Case B
\hei\ emissivities of \citet{Porter:2005} at the respective \elt\ and \eld\ of
the broad and narrow emission components. Since these are only valid up to
log(\eld/cm$^{-3}$) $=$ 6, a correction factor of 0.9543 was applied to the
broad component emissivity to extrapolate up to log(\eld/cm$^{-3}$) $=$ 7.25. At
these densities, collisional excitation from the He$^0$ 2s$^3$S metastable
level by electron impacts contributes to the observed broad \hei\ $\lambda$5876
component; this effect is accounted for by \citet{Porter:2005}.
Fig\,~\ref{fig:He_abundcuts} shows $X$ and $Y$ direction cuts across the broad
and narrow component elemental He/H maps.  We have assumed that the
He$^{2+}$/H$^+$ fraction is negligible. Good agreement between the spatial
variations in the He/H abundance for each component is observed. No relative
He/H enrichment is seen across the components; within the uncertainties both
fall between the pre-galactic He/H ratio of 0.08 \citep[][]{Luridiana:2003} and the solar ratio 0.10 \citep[][]{Lodders:2003}. This
contrasts with the localized N/H enrichment north-east from the nucleus that
spatially correlates with a peak in $EW$(H$\beta$) (see Section~\ref{sec:WR}).
The slight decrease in He/H in the core region might be attributable to the
presence of He$^{2+}$. We find this unlikely, however, since no nebular \heii\
$\lambda$4686 emission has been detected (none was reported in the recent study
by \citet{Thuan:2008}), and in addition, other similarly high excitation lines
such as [Ar\four] are not present in the spectrum of the galaxy. The derived
He/H abundances at the centre of Mrk~996 agree well with those of TIL96 who
calculated He$^+$/H$^+$$\sim$0.09 from the \hei\ $\lambda$5876 emission within
their FOS aperture.

Table~\ref{tab:abundSummary} summarises the derived abundance for the narrow
and broad line emission regions of the galaxy.

\begin{table}
\caption{Adopted abundances for the narrow (C1) and broad (C2) component
emission regions throughout Mrk~996.}
\begin{center}
\begin{minipage}{6cm}
\begin{tabular}{lcc}
\hline
& C1 & C2 \\
Adopted \elt\ (K) & 10,000$^a$ & 11,000\\
Adopted \eld\ (cm$^{-3}$) & 170& 10$^7$\\
\hline
He/H & 0.091$\pm$0.017 & 0.072$\pm$0.018\\
N$^+$/H$^+$ $\times$10$^5$ &0.46$\pm$0.18 & 4.1$\pm$1.1\\
N$^{2+}$/H$^+$ $\times$10$^4$ & -- &1.3$\pm$0.5 \\
N/H $\times$10$^5$ & 0.89$\pm$0.25&  17.0$\pm$7.6$^b$\\
O$^+$/H$^+$ $\times$10$^4$& 1.3$\pm$0.4  & -- \\
O$^{2+}$/H$^+$ $\times$10$^4$& 1.1$\pm$ 0.2 &  2.0$\pm$0.2\\
O/H $\times$10$^4$& 2.4$\pm$0.4 &  2.3$\pm$1.1$^b$\\
S$^+$/H$^+$ $\times$10$^7$& 10.3$\pm$1.8  &(9.7$\pm$1.8) \\
S$^{2+}$/H$^+$ $\times$10$^6$ & 4.7$\pm$0.9& 5.4$\pm$1.3\\
S/H $\times$10$^6$ & 5.9$\pm$1.6 & 6.6$\pm$1.6 \\
Ar$^{2+}$/H$^+$  $\times$10$^7$  & 8.19$\pm$1.8 & 12.0$\pm$2.8 \\
Ar/H  $\times$10$^6 $ &1.5$\pm$ 0.3 & 2.2$\pm$ 0.5 \\
log N/O & $-$1.43 & $-$0.13 \\
log S/O & $-$1.61 & $-$1.54 \\
log Ar/O & $-$2.20 & $-$2.02 \\

\hline
\end{tabular}
\begin{description}
\item[$^a$] Upper limit adopted \elt\ so that C1 abundances are lower limits.
\item[$^b$]  Applicable to a central 0.$''$86 (FOS) aperture; see the text for
details.
\end{description}

\end{minipage}
\end{center}\label{tab:abundSummary}
\end{table}

\begin{figure}
\begin{center}
\includegraphics[scale=0.5]{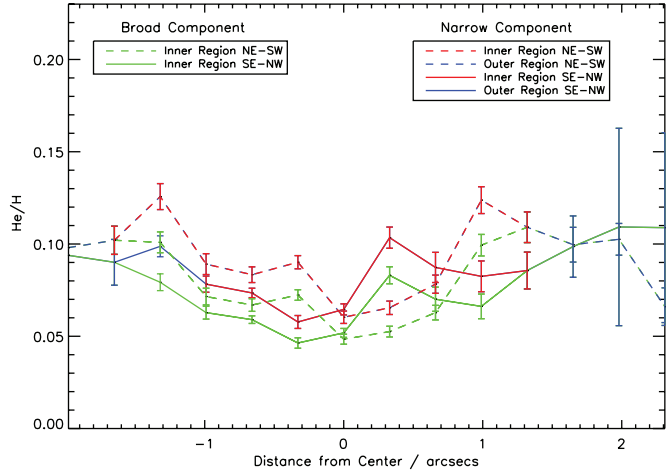}
\caption{Spatial variations in the narrow (blue and red) and broad (green)
component He/H elemental abundance across Mrk~996. Three spaxel-wide averages
were made across the IFU aperture in its $X$ and $Y$ directions, corresponding
to the south-east to north-west and north-east to south-west, respectively.}
\label{fig:He_abundcuts}
\end{center}
\end{figure}

\section{Wolf-Rayet Stars and Stellar Age}
\label{sec:WR}
\begin{figure}
\begin{center}
\includegraphics[height=7.5cm,width=7.5cm]{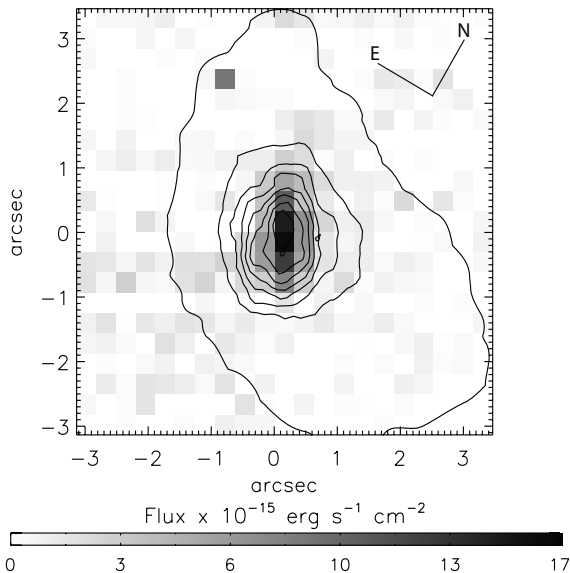}
\caption{Emission map of the blue WR feature (created by integrating over the
full emission feature and removing contaminant flux from [Fe\three]
$\lambda$4658), showing the distribution of WR stars in the inner core region of
Mrk~996. Overlaid are the H$\alpha$ narrow component flux contours from
Fig\,~\ref{fig:HSToverlay}.} \label{fig:WRmap}
\end{center}
\end{figure}

Wolf-Rayet (WR) features were identified in Mrk~996 by TIL96 in their single
aperture FOS spectroscopy of a 0.$''$86 region at the centre of Mrk~996. The
VIMOS IFU spectra also show a broad Wolf-Rayet (WR) stellar feature at
4650\,\AA, attributable to a mixture of late-type WN (WNL) and WC stars at
\heii\ $\lambda$4686 and \niii\ $\lambda$4640 (with an additional weak WC
stellar \civ\ feature at 5808\,\AA). The WR spectral signatures are seen to
extend throughout the core region of Mrk~996 and show excellent agreement with
the spatial extent of the broad \foiii\ $\lambda$4363 emission
(Fig\,~\ref{fig:O3cregions}). A map of the 4650\,\AA\ WR feature is shown in
Fig\,~\ref{fig:WRmap}; this has been decontaminated from the emission by
nebular \ffeiii\ $\lambda$4658 by fitting a Gaussian profile to the line and
subtracting its flux from the integrated flux of the 4650\,\AA\ feature. After
summing the spectra over the core region, LMC WR spectral templates from
\citet{crowther:2006} were used to fit both spectral features. Although the WC
feature is weak, its contribution is essential in fitting the broad wings of
the 4650\,\AA\ feature.  At the distance of Mrk~996 (22~Mpc) the two WR
features indicate the presence of $\sim$2600 WNL stars and $\sim$400 WC stars.
Assuming a similar depth for the area over which these spectral features are
summed, 1.7$\times$2.3 arcsec$^{2}$ (4.6$\times$10$^4$ pc$^2$), results in a
volume density of 3$\times$10$^{-4}$ WR stars per pc$^3$ in the core region.

We can compare our WR number estimates with those of TIL96, who estimated that
600 late-type WN stars and 74 WC stars were responsible for the WR emission
seen within their 0.$''$86 diameter FOS aperture. Scaling our own core region
WR number estimates for the difference in aperture size would predict 390 WNL
stars and 60 WC stars within the FOS aperture.  While the agreement between the
two estimates is quite good, the decrease in WNL star numbers for our scaled
estimate can be attributed to the non uniform density of WR stars in the core
region (see Fig\,~\ref{fig:WRmap}). In contrast, the WC 5808\,\AA\ feature is
very centrally peaked, thus giving a good agreement between the two estimates.

Luminosities of hydrogen recombination lines, particularly H$\beta$, can
provide estimates of the ionizing flux present, assuming a radiation-bounded
nebula \citep[][]{Schaerer:1998}.  Thus, the equivalent width ($EW$) of
H$\beta$ is commonly used as a stellar age indicator at a given metallicity. A
map of $EW$(H$\beta$) across Mrk~996 (derived from integrating over the full
emission line profile) is shown in Fig\,~\ref{fig:HbetaEW} and displays a
surprising morphology compared to the central location of the ionizing sources
in Mrk~996. A peak is seen directly NE of the centre of the galaxy, which
correlates well with the peak in [N\two] $\lambda$6584 emission shown in
Fig\,~\ref{fig:N_H_abund}.  We can use the $EW$(H$\beta$) map, in conjunction
with the metallicity map (described in Section~\ref{sec:abund_nar}), to
estimate the distribution of stellar ages throughout Mrk~996.  We find that the
core region $EW$(H$\beta$) corresponds to a stellar age of 4.5\,Myr whereas the
$EW$(H$\beta$) peak NE of the core corresponds to a slightly younger stellar
age of $\sim$3\,Myr \citep[cf. fig\,~7 of ][]{Schaerer:1998}; these ages are
sufficiently long to allow for the presence of WR stars in the galaxy.

An estimation of the number of O-type stars was made, assuming that all
ionizing photons, $Q^{\rm obs}_{\rm 0}$, are produced by O and WR stars and
utilising the H$\beta$ luminosity, $L$(H$\beta$), integrated over the whole
galaxy (i.e. using the dereddened combined C1 and C2 H$\beta$ flux from the
inner and outer regions, and at a distance of 22.3~Mpc). The absolute number of
O stars was derived using (e.g. \citet{Fernandes:2004})

\begin{equation}
N_{\rm O} = N_{\rm OV} = \frac{Q^{\rm obs}_0 - N_{\rm WR}Q_{\rm WR}}{\eta_{\rm
0}(t)Q_{\rm O7V}},
\end{equation}

\noindent where $Q^{\rm obs}_{\rm 0}$ is the emission rate of ionising photons,
as derived from $L$(H$\beta$), and $Q_{\rm WR}$, $Q_{\rm O7V}$ are the emission
rates of ionizing photons by WR stars and O7V stars, respectively, with $N_{\rm
WR}$ being the total number of WR stars.  $\eta_{\rm 0}(t)$ is a conversion
parameter for the proportion of O7V stars relative to all OV stars and is
estimated as being $\geq$0.2 by \citet{Schaerer:1998} at the metallicity of
Mrk~996 ($\sim$0.5\,\Zsol). Adopting $\eta_{\rm 0}(t)$ $=$0.2, $Q_{\rm WR} =
Q_{\rm O7V} = 1.0\times10^{49}$~s$^{-1}$ \citep[][]{Schaerer:1999}, and $Q^{\rm
obs}_{\rm 0}$(C1+C2)$=$3.36$\times$10$^{53}$ photons s$^{-1}$ we find $N_{\rm
O} \lesssim$ 153,000 yielding $N_{\rm WR}/N_{\rm OV} \gtrsim 0.02$. As a lower
limit, this ratio is in good agreement with those predicted by evolutionary
synthesis models by previous studies; at the predicted age of $<$5~Myr for the
Mrk~996 starburst, \citet{Cervino:1994} find that this $N_{\rm WR}/N_{\rm OV}$
fraction is typical at metallicities of 0.4\,\Zsol, while \citet{Schaerer:1998}
predict $N_{\rm WR}$/($N_{\rm WR}+N_{\rm O}$) $=$ 0.04 for a metallicity of
0.25\,\Zsol.

Using the H$\alpha$--SFR (star formation rate) relation of
\citet{Kennicutt:1998} and the total H$\alpha$ luminosity (C1+C2) of 4.83
$\times$10$^{41}$~erg~s$^{-1}$ we obtain a global current SFR of
3.8\,M$_\odot$~yr$^{-1}$. Taking into account the subsolar metallicity of
Mrk~996 this can be reduced to 2.7\,M$_\odot$~yr$^{-1}$ \citep[][]{Lee:2002}.
However, in the event that the C2 component emission is partly attributable to
shock-excited gas (see Section \ref{sec:narrow_origin}) so that it cannot be
included in the estimate of $Q^{\rm obs}_{\rm 0}$, a SFR lower limit of
1.3\,M$_\odot$~yr$^{-1}$ is derived based on Q$_0$(C1) $=$
1.31$\times$10$^{52}$ photons s$^{-1}$ alone.

\begin{figure}
\begin{center}
\includegraphics[scale=0.45]{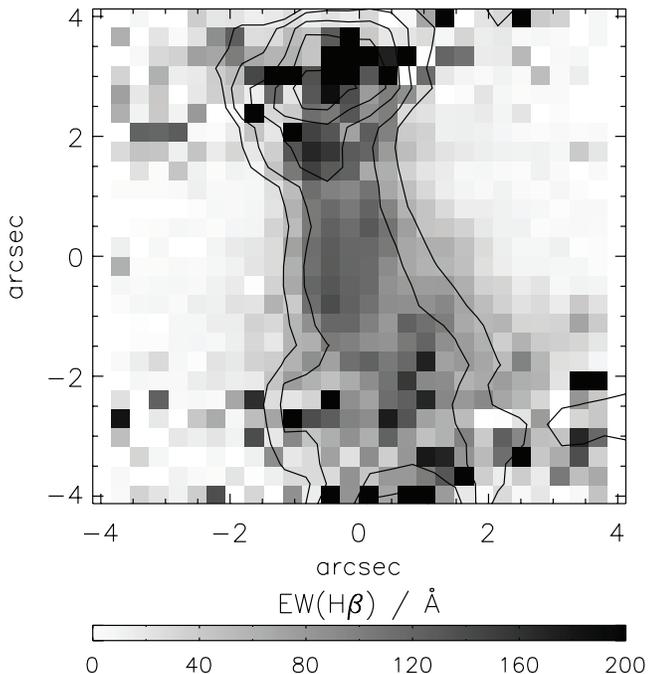}
\caption{Map of the equivalent width of H$\beta$ across Mrk~996, derived from
the full emission line profile.  Contours range from 70--180\,\AA\ in steps of
30\,\AA.} \label{fig:HbetaEW}
\end{center}
\end{figure}

\section{Discussion}
\label{sec:discussion}

\subsection{Diagnosing the ionisation mechanisms}
\label{sec:narrow_origin}

In order to gain insight into the mechanisms responsible for the
multi-component line emission from Mrk~996, we made use of the classic
diagnostic diagrams of \citet{Baldwin:1981} (the BPT diagrams).
These are employed to classify galaxies according to the dominant excitation
mechanism of their emission lines, i.e. either photoionisation by massive stars
within \hii\ regions or photoionisation by non-thermal continua from active
galactic nuclei (AGN). The diagrams consist of excitation-dependent,
extinction-independent line ratios: log([O\three] $\lambda$5007/H$\beta$)
versus either log([N\two] $\lambda$6584/H$\alpha$) or
log([S\two]($\lambda$6716,31)/H$\alpha$). Star-forming galaxies fall into the
lower left region of the diagram, AGN host galaxies fall into the upper right
region and Low-ionisation Emission Line Regions (LINERs) fall in the lower
right region. The separation is not as clear, however, for low metallicity AGNs
(see e.g. \citet{Stasinska:2006}), only a handful of which have been proposed
to exist thus far \citep[][]{Izotov:2008}.

\citet{Kewley:2001} calculated the first starburst grids that attempted to
match optical diagnostic diagrams based on purely empirical data, by coupling
the {\sc mappings iii} photoionisation code and {\sc starburst99} population
synthesis models.  Whereas previous semi-empirical studies involved only
solar-metallicity calculations, their line ratios were computed for a range of
metallicities ($Z$ $=$ 0.05--3.0\,\Zsol) and ionisation parameters ($q$ $=$
$5\times10^6$--$3\times10^8$~cm s$^{-1}$), where $q$ is the maximum velocity of
an ionisation front that can be driven by the local radiation field and is
related to the non-dimensional ionisation parameter $U$ $=$ $q$/c,
\citep[][]{Dopita:2001}. Utilizing these grids, \citet{Kewley:2001} determined
a `maximum starburst line' (shown in Fig\,~\ref{fig:BPT}), above which the flux
ratios of an object cannot be fitted by pure starburst models alone. Ratios
lying above this photoionisation boundary require additional sources of
excitation such as shocks or AGNs. As mentioned above, this picture has
recently been revised and it is now believed that low-$Z$ AGNs, should they
exist, would occupy similar regions on the BPT diagram as normal \hii\
galaxies; based on an analysis of a large sample of Sloan Digital Sky Survey
galaxies, \citet{Stasinska:2006} have found that composite AGN/\hii\ region
models with $Z$ $<$ 0.4\,\Zsol\ lie very close to and below the pure \hii\
region sequence, whereas when considering also the regime of $Z$ $>$
0.6\,\Zsol\ they established that the locus below the \citet{Kewley:2001} line
allows for an AGN-excited gas component of up to 20 per cent.

It is therefore instructive to see what area of the BPT diagram Mrk~996
occupies. \citet{Kewley:2001} found that since the [S\two](6716,31)/H$\alpha$
ratio is affected by the density of the line emitting region and it is
therefore less accurate as a diagnostic. Hence considering the large density
difference between the narrow and broad component gas within Mrk~996 we only
considered the [O\three] $\lambda$5007/H$\beta$ vs. [N\two]
$\lambda$6584/H$\alpha$ ratios (Fig\,~\ref{fig:BPT}).  Since no broad component
is detected for [N\two] $\lambda$6584 we have not been able to create separate
broad and narrow diagnostic diagrams and instead present (i) a narrow component
diagram (Fig\,~\ref{fig:BPT} top panel), and (ii) ratios derived from the
entire emission line profiles (i.e. integrated over broad and narrow component
emission, where both exist; Fig\,~\ref{fig:BPT} bottom panel). Spaxels
corresponding to the core region of the galaxy are shown as red crosses, with
outer region spaxels plotted as green crosses. Also shown is the maximum
starburst line of \citet{Kewley:2001}.

The data points for the outer narrow line region of Mrk~996 mostly straddle the
`theoretical' upper limit for pure photoionisation. Based on the
\citet{Kewley:2001} grids, the flux ratios lie within a metallicity range of
$Z$ $=$ 0.3--0.5 and along and above an ionisation parameter track for $q$ $=$
3$\times10^8$ (the maximum ionisation parameter computed). The narrow component
emission in the core region occupies the top left of the distribution, which
corresponds to a higher ionisation parameter and lower metallicity in the
starburst grids, while a significant number of data points fall outside the SFG
(star-forming galaxies) area of the plot. When the broad component is included
in the diagnostics, the location of the outer region spaxels move to the left,
whereas the core region spaxels move downwards and to the left for both ratios
to occupy a lower ionisation parameter and metallicity region.  This is most
likely due to the suppression of the [O\three] $\lambda$5007 broad component
emission within the high density core region, which would lower the combined
narrow plus broad [O\three] $\lambda$5007/H$\beta$ ratio.

As previously mentioned, exceeding the `maximum starburst' line can also be
indicative of a contribution to the ionisation by fast shocks. However, the
amount by which this line can be crossed before shocks must be a substantial
source of ionisation has not been established thus far. Similar diagrams were
created by \citet{Calzetti:2004} for four starburst (SB) galaxies using ratios
from {\it HST} WFPC2 images, such as [O\three] $\lambda$5007/H$\beta$ vs.
[S\two] $\lambda$6716,31/H$\alpha$. They found that a number of regions in the
SB galaxies lay above the maximum starburst line and as a result defined
regions of non-stellar excitation as either lying above and to the right of it
or where the [S\two]/H$\alpha$ ratio is high enough to be compatible with
non-stellar sources, i.e. [S\two]/H$\alpha$ $>$ 0.7--0.9
\citep[][]{Veilleux:1987,Shull:1979}.  The log([S\two]
$\lambda$6716,31/H$\alpha$) ratio is traditionally used as a diagnostic for
shock-excited gas because shock models predict that relatively cool
high-density regions should form behind the shock front and emit strongly in
[S\two] \citep[][]{Dopita:1978}. The [S\two]/H$\alpha$ flux ratios for the
narrow component emission across Mrk~996 lie between 0.03--6$\times10^{-4}$,
well below the high [S\two]/H$\alpha$ ratios predicted for shocks.  On the
other hand, \citet{Thuan:2008} employed a model with shocks at a velocity of
250 km~s$^{-1}$ that reproduced the observed intensity of the high-ionisation
potential [O~{\sc iv}] 25.9\,$\mu$m line detected in the {\it Spitzer} spectrum
of Mrk~996.  In conclusion, Fig\,~\ref{fig:BPT} by itself cannot provide firm
evidence for or against the presence of AGN activity within Mrk~996 given the
galaxy's rather low metallicity status, even though the integrated broad
component H$\alpha$ luminosity of 2.5$\times$10$^{41}$~erg~s$^{-1}$ this at the
lower limit of those measured from rare low-Z BCDs suspected of harbouring active galactic nuclei
\citep[3$\times$10$^{41}$ -- 2$\times$10$^{42}$~erg~s$^{-1}$ ][]{Izotov:2008}. Although photoionisation appears to be the dominant
excitation mechanism, we cannot rule out a contribution from shocks,
particularly with respect to the broad component emission.  We draw attention
to the significant bias inherent in BPT diagrams of galaxies whose analysis
does not involve a separate consideration of narrow/broad line diagnostics.

\begin{figure*}
\begin{center}
\includegraphics[scale=0.6]{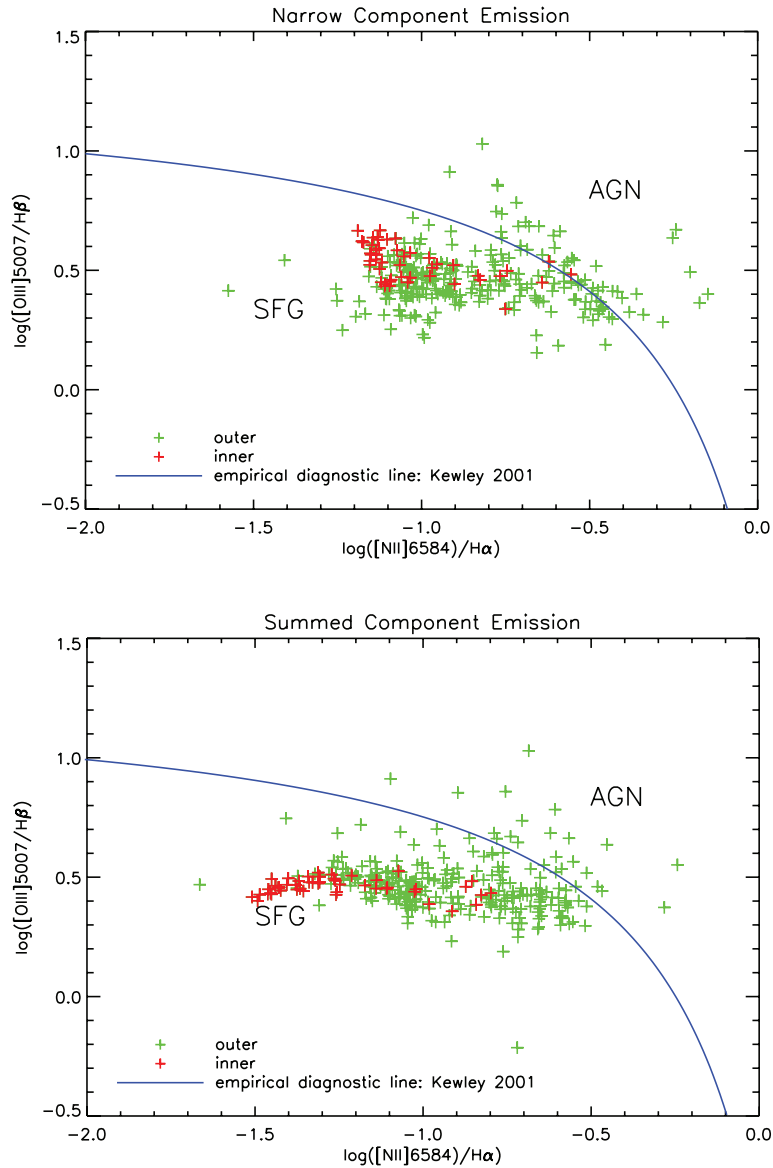}

\caption{Emission line diagnostic diagram for Mrk~996.  Each data point
represents a spaxel in the ratioed dereddened flux maps corresponding to
[O\three] $\lambda$5007/H$\beta$ and [N\two] $\lambda$6584/H$\alpha$.  The
empirical diagnostic `maximum starburst' line from \citet{Kewley:2001} is also
shown, as are the positions of emission line ratios for star-forming galaxies
(SFG) and active galactic nuclei (AGN).  The top panel represents narrow
component emission line map ratios whereas the bottom panel shows ratios
derived from the full emission line profile (i.e. narrow $+$ broad component
emission, where detected).} \label{fig:BPT}
\end{center}
\end{figure*}

\subsection{What is the origin of the broad line emission?}
\label{sec:broad_origin}

Mrk 996 is not the only BCD that shows evidence for large line widths.
\citet{Izotov:2007} presented an analysis of a large number of low metallicity
BCDs exhibiting broad emission lines with inferred gas expansion velocities of
1000--7000 km/s. WR stellar winds have been proposed as a possible mechanism
for producing such broad line widths, but thus far no correlation between the
presence of broad nebular lines and broad WR features have been found
\citep[][]{Izotov:2007}. We find that in Mrk~996 the surface brightness of the
4650\AA\ WR feature does not correlate tightly with that of \foiii\
$\lambda$4363 (Fig\,~23), further indicating that the mechanism responsible for
the broad line emission is not active only in the regions where WR stars are
present.

 \citet{Roy:1991,Roy:1992} investigated a variety of mechanisms
to explain broad nebular gas components in \hii\ regions, including 
electron scattering, stellar winds, supernova remnants and superbubble 
blowouts.  However, each of these mechanisms was deemed unsatisfactory.  
An exploration of blowout mechanisms within low-metallicity \hii\ regions 
was conducted by \citet{Tenorio-tagle:1997} using a hydrodynamical 
calculations to try to match a sample of low-metallicity \hii\ regions 
with broad emission line components.  They proposed that the absence 
of strong radiative cooling in the low-metallicity ISM within these 
regions could delay the action of  Rayleigh-Taylor (RT) instabilities 
that can fragment the expanding shell within a blowout.  However, this 
mechanism may not be applicable to Mrk~996, as one of the required 
conditions is a low-density environment, whereas we derive high densities 
for its broad component gas (log(\eld\ )$\sim$7).

Broad line emission (FWHM $\approx$ 400 km/s; C2) is detected throughout the
central regions of Mrk 996, throughout the bright starburst region and out to
$>$5$''$. Close to the nucleus of the galaxy the broad Balmer line component
splits into two additional \emph{broad} components (FWHM $\approx$
200--300\,km~s$^{-1}$) which trace the kinematics of a two-armed mini spiral,
but component C2 does not show evidence of velocity splitting at our level of
spectral resolution farther out (Figs\,~3 and 4). Given the youth of the
starburst region in Mrk~996 ($<$5~Myr) containing large numbers of young O and
WR-type stars (Section 6), and presence of very dense zones of gas (Section
4.3), it is possible that the broad line emission partly originates from a
turbulent mixing layer forming on the surface of dense cool clumps which are
subjected to irradiation and hydrodynamic ablation from the hot winds from
young massive star clusters (\citet{Begelman:1990}; \citet{Slavin:1993}).
Recent work on the nearby NGC 1569 and M82 has revealed that the broad line
component observed in those starburst galaxies can be explained as originating
within turbulent layers on the surface of the dense gas clumps \citep[][]{Westmoquette:2007a,Westmoquette:2007b}.  Motivated by this hypothesis, \citet{Binette:2009} have successfully modelled the broad underlying components seen in the Balmer and \foiii\ lines of NGC~2363 via the  inclusion of a turbulent mixing layer and conclude that the broad profile results from radial acceleration of photoionised turbulent gas.  Their models also correctly predict the absence of a broad component in \fsii\ and \fnii\ lines.  Although the applicability of these results to Mrk~996 may be limited due to the high densities of Mrk~996's broad components (\eld\ $=$\,100\,cm$^{-3}$ is adopted for NGC~2363) and their much broader predicted line widths ($\pm$3500\,kms$^{-1}$), their results lend a great deal of weight to the turbulent mixing layer hypothesis.

This mechanism could also be at work here. It is
probable that the central region of Mrk 996 consists of many small, dense
clouds of relatively low volume filling factor dispersed within the young star
clusters of the starburst. This results in a large cloud surface area with
which the copious ionizing photons and fast stellar winds can interact. If the
presence of broad emission indicates that strong wind-clump interactions are
taking place, then by extension, material from these interaction sites must be
being stripped off and entrained into the cluster wind flows contributing to
the overall appearance of component C2.

\begin{figure*}
\begin{center}
\includegraphics[scale=0.6]{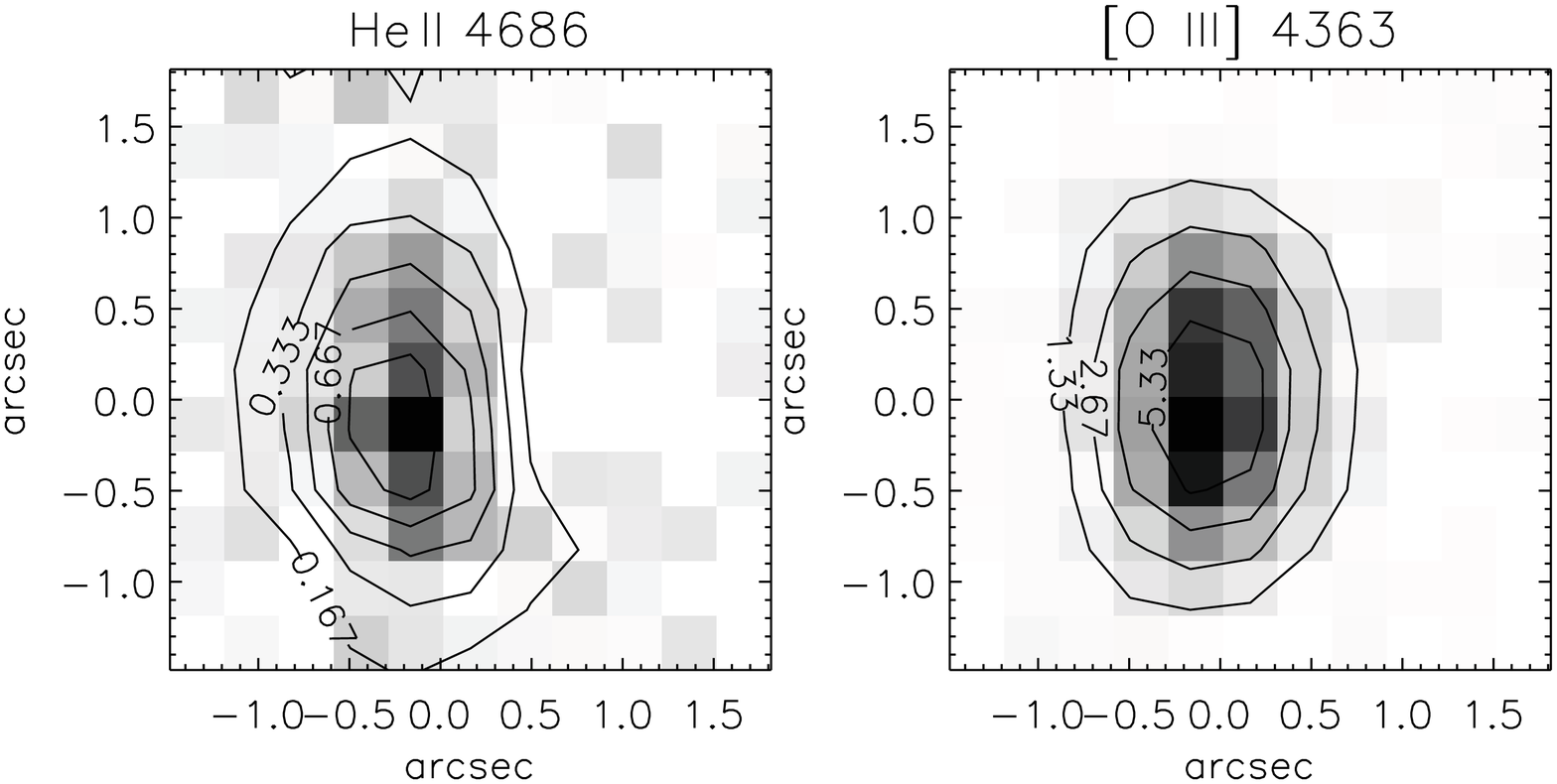}

\caption{Grey scale and contour emission line maps of the He\two~$\lambda$4686
WR feature (left panel) and nebular [O\three]4363 (right panel).  Their
contours suggest that the ionisation mechanism responsible for the broad
emission line components (represented by [O\three]~$\lambda$4363) may not be
strictly due to the WR stars. Flux contours are in units of $\times10^{-15}$erg
s$^{-1}$ cm$^{-2}$ per 0.33$\times$0.33 arcsec spaxel.} \label{fig:HeI4684}
\end{center}
\end{figure*}

\section{Summary and conclusions}

Mrk~996 is a complex system containing an extended starburst region
(216--432~pc in radius) with a centrally concentrated Wolf-Rayet population in
the nuclear regions ($\lesssim$200~pc). The young stellar population in the
nuclear super-cluster contains $\sim$3000 WR stars and $\lesssim$150,000
O7V-type equivalent stars. The age of the nuclear starburst is estimated to be $\sim$4.5~Myr
but there is evidence that the extended starburst is younger by $\sim$1~Myr.
The presence of an old stellar population ($\gtrsim$1~Gyr) has been
independently established (Thuan, Hunt \& Izotov 2008). The current star
formation rate is $\sim$2\,M$_\odot$\,yr$^{-1}$. The spatially mapped ionized
gas shows a composite emission spectrum consisting of narrow and broad lines.
The Balmer line velocity structure close to the nucleus confirms the presence
of a spiral structure confined in the inner 160~pc. The H$\alpha$ broad
component has a large integrated luminosity of
2.5$\times$10$^{41}$~erg~s$^{-1}$; this is at the lower limit of those measured
from the rare low-$Z$ BCDs suspected of harbouring active galactic nuclei \citep[][]{Izotov:2008}. Our
VIMOS IFU analysis has enabled a separate analysis of the physical conditions
and chemical composition of the narrow and broad emission line regions. The
broad line region in the nuclear starburst is very dense (10$^7$\,cm$^{-3}$)
whereas the narrow line component is of lower density
($\lesssim$10$^3$\,cm$^{-3}$).

The upwards revised oxygen metallicity of Mrk~996 is $\geq0.5$\,\Zsol\ (12 $+$ log
O/H~$\approx$~8.37). The broad line region is nitrogen-enriched compared to the
narrow line region by $\lesssim$1.3 dex. However, no relative enrichment of He,
O, S, and Ar is inferred. The S/O and Ar/O abundance ratios in the two
components, as well as the N/O ratio in the \emph{narrow} line component, are
typical of those in \hii\ galaxies and dwarf irregulars
\citep[e.g.][]{VanZee:2006,Izotov:2006}. The narrow-line N/O ratio in
particular is exactly the value expected based on the galaxy's colour, and
follows the metallicity-luminosity relationship for isolated dwarf galaxies
\citep[cf. figs\,~8, 9 of ][]{VanZee:2006} for $M_B$ $=$ $-$16.7, $B-V$ $=$
0.44, TIL96); its mean current value can be interpreted as the result of the slow release of
nitrogen from the intermediate-mass stellar population over the last few Gyrs.
On the other hand, the high N/O ratio in the broad line region of the inner
galaxy is consistent with the presence of numerous evolved massive stars (e.g.
WNL-type and Luminous Blue Variables) and can be attributed to the cumulative
effect of their N-enriched winds. An elevated N/O could not in this case be due
to the outflow of oxygen-enriched gas from supernova explosions: the absence of
elevated (or suppressed) O/H, S/H and Ar/H ratios in the broad line gas implies
that recent supernova ejecta are probably not implicated in its excitation; at
the young age of the Mrk~996 starburst only a few very massive stars
($\gtrsim$50\,M$_\odot$) would have exploded as supernovae \citep[e.g.
][]{Woosley:2002}. The mild two-fold increase over the mean of the narrow-line
N/H (Fig\,~\ref{fig:He_abundcuts}) spatially correlates with a local peak in
$EW$(H$\beta$) in a region (Fig\,~\ref{fig:HbetaEW}) where no WR stars are
seen. This could mean that even though the narrow-line gas is fairly well
mixed, some localized N-enrichment has already occurred in an area dominated by
normal OB-type stars which is slightly younger than the nuclear starburst.

Finally, we would like to draw attention to the nature of the broad \foiii\
$\lambda$4363 line detected in Mrk~996. Even though this galaxy represents an
extreme case in harbouring a very dense ionized component which dominates the
excitation of this line, it is by no means unique \citep[see e.g.
][]{Izotov:2007}. In such cases where a substantial
fraction of the \foiii\ 4363 and 4959, 5007\,\AA\ lines arise in different gas
components, the integrated auroral to nebular ratio would \emph{not} be
representative of the electron temperature of the overall \hii\ region; using it would result in biased physical conditions and chemical abundances. The problem
would be aggravated in analyses based on low dispersion, low spatial resolution
spectra and may therefore have implications for the so-called `heating
problem' of blue compact galaxies
\citep[e.g.][]{Stasinska:1999,Pequignot:2008}.

\section{Acknowledgments}

We thank the VIMOS support staff at Paranal for taking these service mode
observations [programme 078.B-0353(A), PI: Tsamis]. We appreciate discussions
with Marina Rejkuba and Carlo Izzo about the VIMOS instrument and the GASGANO
reduction pipeline. Also, our thanks go to Fabrizio Sidoli 
for helpful discussions regarding WR template
spectra and to Nate Bastian for advice on IFU cube construction. This 
research has made use of the NASA ADS database. BLJ acknowledges
support from a STFC studentship. YGT acknowledges support from a STFC
fellowship and from grants AYA2007-67965-C03-02 and CSD2006-00070
CONSOLIDER-2010 ``First science with the GTC" of the Spanish Ministry of
Science and Innovation.


\bibliography{}

\appendix

\bsp

\label{lastpage}

\end{document}